\newif\iflncs\lncsfalse
\newif\ifmai\maifalse
\newif\ifanon\anonfalse
\documentclass{lmcs}
\pdfoutput=1

\usepackage{lastpage}
\lmcsdoi{15}{1}{14}
\lmcsheading{}{\pageref{LastPage}}{}{}%
{Oct.~24,~2017}{Feb.~15,~2019}{}

\usepackage{etex}
\usepackage[utf8]{inputenc}
\usepackage{amsmath}
\usepackage{url,amsmath,amsfonts}

\usepackage{gastex}

\newif\ifnoblind\noblindfalse
\usepackage{aecompl}
\usepackage[mathscr]{euscript}
\usepackage{graphics}
\usepackage{graphicx}

\usepackage{pstricks}
\usepackage{stackengine}
\usepackage[mathscr]{euscript}%
\usepackage[english]{babel}
\usepackage[all]{xy}
\usepackage{url}
\newcommand{\goto}[1]{\ensuremath{ \xrightarrow{\;#1\;} }}

\usepackage{amssymb,amsmath}
\usepackage{wasysym}
\usepackage{pgf}
\usepackage{tikz}
\usepackage{wrapfig}
\usepackage{epstopdf}

\newenvironment{proof_of}[1]{\leavevmode\begin{proof}[Proof of #1.]}{\end{proof}}
\newcommand{\defin}{\stackrel{\triangle}{=}}%{\bigtriangleup}{=}}

\newcommand{\vspm}{\vspace*{-.4cm}}

\newcommand{\reals}{{\mathbb{R}}}

\newcommand{\ode}{{\sc ode}}
\newcommand{\bde}{{\sc bde}}
\newcommand{\fde}{{\sc fde}}
\newcommand{\mor}{{\sc mor}}
\newcommand{\vecx}{\mathrm{\mathbf{x}}}
\newcommand{\xx}{\mathrm{\mathbf{x}}}
\newcommand{\yy}{\mathrm{\mathbf{y}}}
\newcommand{\zz}{\mathrm{\mathbf{z}}}

\newcommand{\aalpha}{ {\mathbf{\phi}}}
\newcommand{\field}{\mathbb{R}}
\newcommand{\lie}{\mathcal{L}}
\newcommand{\ide}[1]{\big\langle\,\, #1 \,\,\big\rangle}%{\left\langle\left\langle #1 \right\rangle\right\rangle}
\newcommand{\A}{\mathcal{A}}
\renewcommand{\aa}{\mathrm{\mathbf{a}}}
\newcommand{\myspan}{\mathrm{span}}
\newcommand{\pro}{\mathrm{pr}}

\newcommand{\Zzero}{\mathcal{Z}_\ivp}
\newcommand{\M}{\mathcal{M}}
\newcommand{\UM}{U_{\M}}
\newcommand{\rk}{\mathrm{rk}}
\newcommand{\ivp}{\mathbf{\Phi}}
\newcommand{\redpr}{\mathbf{\Psi}}

\newcommand{\lbisim}{$\lie$}

\usepackage{scalerel}
\stackMath
\newcommand\bighat[1]{%
\savestack{\tmpbox}{\stretchto{%
  \scaleto{%
    \scalerel*[\widthof{\ensuremath{#1}}]{\kern-.6pt\bigwedge\kern-.6pt}%
    {\rule[-\textheight/2]{1ex}{\textheight}}%WIDTH-LIMITED BIG WEDGE
  }{\textheight}%
}{0.5ex}}%
\stackon[1pt]{#1}{\tmpbox}%
}


\begin{document}

%%%%%%%%%%%%%%%%%%%%%%%%%%%%%%%%%%%%%%%%%%%%%%%%%%%%%%%%%%%%%%%%%%%%%%%%%%%%%%
%%%%%%%%%%%%%%%%%%%%%%%%%%%%%% Title %%%%%%%%%%%%%%%%%%%%%%%%%%%%%%%%%

\title[Coalgebra and differential equations]{Algebra, coalgebra,  and minimization\\
 in polynomial differential equations}
\titlecomment{{\lsuper*} Extended and revised version of \cite{Bor17}.}

\author[M.~Boreale]{Michele Boreale}
\address{Universit\`a di Firenze,
Dipartimento di
Statistica, Informatica, Applicazioni (DiSIA) ``G. Parenti'', Viale Morgagni 65, I-50134
 Firenze, Italy.}
\email{michele.boreale@unifi.it}  %optional

%\date{}

\begin{abstract}
We consider reasoning  and minimization in systems
of polynomial ordinary
 differential equations (\ode's). The ring of  multivariate polynomials is employed as a
 syntax for denoting system behaviours.
  We endow this set  with a transition system structure based on the   concept of \emph{Lie
  derivative}, thus inducing a notion of  \lbisim-\emph{bisimulation}. We prove that two states (variables) are   \lbisim-bisimilar
  if and only if they correspond to the same  solution
  in the \ode's system. We then characterize \lbisim-bisimilarity algebraically,
   in terms of certain   ideals in the polynomial ring that are invariant under Lie-derivation.
  This characterization allows us to develop a complete algorithm, based on
  building an ascending chain of ideals, for computing the largest
   \lbisim-bisimulation
  containing all valid identities that are instances of a user-specified template. A specific largest
  \lbisim-bisimulation
   can be used to build a
  reduced system of \ode's, equivalent to the original one, but
  \emph{minimal}  among all those obtainable by linear aggregation of the original
    equations.
   A  computationally less demanding {approximate}
   reduction and linearization technique is also proposed.
\end{abstract}

\maketitle

\section{Introduction}\label{sec:intro}
%We consider some algebraic and coalgebraic aspects of systems
%defined by  polynomial ordinary differential equations (\pode).
%In many branches of Science, systems of ordinary differential
%equations (\ode's) are the  language used to describe  the laws
%governing evolution of systems through time.
The past few years have
witnessed a surge of interest in computational models based on
 ordinary differential
equations (\ode's), ranging from continuous-time Markov chains (e.g.
\cite{CTMC}) and process description languages oriented to
bio-chemical systems (e.g. \cite{RifBio1,RifBio2,RifBio3}), to
deterministic approximations of stochastic systems (e.g.
\cite{RifBio4,Fluid}), and   hybrid systems (e.g.
\cite{Tiwa,San04,DDL}).

From a computational point of view, our motivation  to study \ode's
arises from the following problems.
\begin{enumerate}
\item \emph{Reasoning}:    provide methods to  {automatically}  prove  and discover  identities
involving the system variables.
\item \emph{Reduction}: provide methods to  {automatically} reduce, and    possibly minimize,
 the number of variables
and equations of a system, in such a way that the reduced system
retains all the relevant information of the original  one.
\end{enumerate}
%Reasoning and reduction are of paramount importance,  conceptually
%and pragmatically.
Reasoning may help an expert  (a chemist, a biologist, an engineer)
to prove or to disprove certain  system properties, even before
actually solving, simulating  or realizing the system. Often, the
 identities of interest    take the form of
\emph{conservation laws}. For instance, chemical reactions often
enjoy a mass conservation law, stating that the sum  of the concentrations of two or more chemical species,
is a constant. More generally, one would like tools to automatically
\emph{discover} all laws of a given form. Pragmatically,
 before actually solving or simulating a given system, it can be critical
  to reduce the system to a size that can be handled by a
 solver or a simulator.

Our goal   is  showing that  these issues can be dealt with  by   a mix of algebraic and
coalgebraic techniques. We will consider \emph{initial value}
problems,  specified by  a system of \ode's of the form $\dot x_i =
f_i(x_1,...,x_N)$, for $i=1,...,N$, plus initial conditions. The
functions $f_i$s are called \emph{drifts}; here we will focus on the
case where the drifts are multivariate polynomials in the variables
$x_1,..,x_N$. Practically, the majority of functions found in
applications is in, or can be encoded into this format (possibly
under restrictions on the initial conditions), including
exponential, trigonometric, logarithmic and rational functions.

A more detailed account of our work follows. We introduce the ring
of multivariate polynomials as a syntax for denoting the
\emph{behaviours} induced by the given initial value problem
(Section \ref{sec:prel}). In other words, a behaviour is any
polynomial combination of the individual components $x_i(t)$
($i=1,..,N$) of the (unique) system solution.
%, but also in any
%polynomial combination of them.
  We then endow the polynomial ring with a transition system, based on a purely syntactic notion  of \emph{Lie
  derivative} (Section \ref{sec:coalg}). This structure naturally  induces a notion of bisimulation over polynomials,
    \lbisim-\emph{bisimilarity}, that is in agreement with the underlying \ode' s.
 In particular, any two variables $x_i$ and $x_j$ are
\lbisim-bisimilar if and only
   the corresponding solutions are the same,  $x_i(t)=x_j(t)$ (this generalizes to polynomial behaviours as expected).
   This way, one can prove identities between two behaviours, for   instance conservation laws, by
   exhibiting  bisimulations containing the given
   pair
   \iflncs
   (in \cite{Full} we show how to   enhance the resulting proof technique by introducing a
   polynomial version of the \emph{up to} techniques  of \cite{San}.
   \else. The resulting proof method is greatly  enhanced by  introducing a
   polynomial version of the \emph{up to} technique  of \cite{San}.
   \fi
 In order to turn this method into a fully automated proof procedure,
 we first  characterize \lbisim-bisimulation  algebraically, in terms of
   certain \emph{ideals}  in the polynomial ring that are invariant under Lie-derivation (Section \ref{sec:bisim}).
 %we first develop a
%characterization   the
 % largest \lbisim-bisimulation in terms of \emph{ideals} over the
 % polynomial ring.
  This characterization leads to an algorithm that, given a user-specified template,
  returns   the set of all its instances that are valid identities in the system
   (Section \ref{sec:algorithm}). One may use this algorithm,   for instance,   to discover all the
   conservation laws  of the system involving terms up to a given degree.
   The algorithm implies building an ascending chain of ideals
   until stabilization, and relies on
   a few basic concepts from Algebraic Geometry, notably Gr\"{o}bner
   bases \cite{Cox}.
  The output of the algorithm is in turn essential to build a
  reduced system of  \ode's, equivalent to the original one,
  but featuring a   \emph{minimal}  number of equations and variables, in the class of systems that can be
  obtained by linear aggregation from the original one (Section
  \ref{sec:minim}).
%\ifmai
   A  computationally less demanding \emph{approximate} reduction and linearization technique is
   proposed (Section \ref{sec:approx}). This may be an attractive
   alternative, because it is entirely based on linear algebraic,
   hence efficient,
   techniques, and  produces `small' linear systems. In particular,  $m$ linear equations
   are sufficient to guarantee an approximation within $O(t^m)$ of the original system.
%\fi
    We then illustrate
   the results of some simple experiments we have conducted using
    a prototype implementation (in Python) of our
   algorithms (Section \ref{sec:examples}). Our approach is mostly related to some recent work on equivalences for \ode's
   by Cardelli et al. \cite{MircoPopl} and to work in the area of hybrid systems.
     We discuss this and other related work, as well as some possible directions for future work, in the concluding section
   (Section \ref{sec:concl}).
   \iflncs
   Due to space limitations, most proofs and   examples are omitted from the present version; they can be found in a technical report available online \cite{Full}.
   \else
   In the interest of readability, some   proofs and    technical
   material have been confined to  a separate appendix (Appendix \ref{app:proofs}).
   %; a few straightforward proofs
   %have been omitted right away from the present version.

To sum up, we give the following contributions.
\begin{enumerate}
\item A complete bisimulation-based proof technique  to reason on
the polynomial behaviours induced by a system of \ode's.
\item An algorithm to find all the valid
 polynomial identities induced by the given system and fitting a
user-specified template.
\item An algorithm to build a reduced,  equivalent system   that is minimal in the class of all
linear aggregations of the original system.
%This can also used to
%decide equivalence of two systems.
\item An algorithm to build a small linear system, approximating within $O(t^m)$ the original system.
\end{enumerate}
\fi

\section{Preliminaries}\label{sec:prel}
Let us fix an integer $N\geq 1$ and a set of $N$ distinct variables
$x_1,...,x_N$. We  will denote by $\vecx$ the
column\footnote{\emph{Vector} means column vector, unless otherwise
specified.} vector $(x_1,...,x_N)^T$.
%(all vectors are columns vectors unless otherwise specified).
We let $\field[\vecx]$ denote the set of multivariate polynomials in
the  variables $x_1,...,x_N$ with coefficients in $\field$, and let
$p,q$ range over it. Here we regard polynomials as syntactic
objects. Given an integer $d\geq 0$, by $\field_d[\vecx]$ we denote
the set of polynomials of degree $\leq d$.  As an example,
$p=2xy^2+(1/5)wz+yz+1$ is a polynomial of degree $\deg(p)=3$, that
is $p\in \field_3[x,y,z,w]$, with monomials $xy^2$, $wz$, $yz$ and
1. Depending on the context, with a slight  abuse of notation it may
be convenient to let a polynomial denote   the induced function
$\field^N \rightarrow
 \field$, defined as expected. In particular, $x_i$ can be seen as denoting the projection on the $i$-th
 coordinate.

A (polynomial) \emph{vector field} is a  %(column)
vector of  $N$ polynomials,  $F=(f_1,...,f_N)^T$, seen as a function
$F:\mathbb{R}^N\rightarrow \mathbb{R}^N$.
A vector field $F$ and an initial condition $v_0\in \field^N$
together  define   an \emph{initial value problem} $\ivp=(F,v_0)$,
often written in the following form
\begin{equation}\label{eq:ivp}
\ivp:\;\left\{\begin{array}{rcl}
{\dot\vecx(t)} & = & F(\vecx(t))\\
\vecx(0) & = & v_0\,.
\end{array}
\right.
\end{equation}
The functions $f_i$  in $F$ are called \emph{drifts} in this
context. A \emph{solution} to this  problem is a  differentiable
function $\vecx(t): D\rightarrow \field^N$,  for some nonempty open
interval $D\subseteq \mathbb{R}$ containing 0, which fulfills the
above two equations, that is: $\frac d{dt}\vecx(t)=F(\vecx(t))$ for
each $t\in D$  and $\vecx(0)=v_0$. By the Picard-Lindel\"{o}f
theorem \cite{PL}, there exists a nonempty open interval $D$
containing 0, over which there is a \emph{unique} solution, say
$\vecx(t)=(x_1(t),...,x_N(t))^T$,  to the problem. In our case,  as
$F$ is infinitely often differentiable,   the solution is seen to be
\emph{analytic} in $D$: each $x_i(t)$  admits a Taylor series
expansion   in a neighborhood of 0. For definiteness, we will take
the domain of definition  $D$ of $\xx(t)$  to be the largest
symmetric open interval  where for each $i=1,...,N$, the Taylor expansion from 0 of    $x_i(t)$ converges pointwise to $x_i(t)$ (possibly $D=\field$). The
resulting vector function of $t$, $\vecx(t)$,  is called the
\emph{time trajectory} of the system.

Given a differentiable function $g:E\rightarrow \mathbb{R}$, for
  some  open set $E\subseteq \mathbb{R}^N$, the \emph{Lie
derivative of $g$ along $F$} is  the function $E\rightarrow
\mathbb{R}$ defined as
\begin{eqnarray*}
\lie_F(g) & \defin & \langle \nabla g, F\rangle = \sum_{i=1}^N
(\frac{\partial g}{\partial x_i} \cdot f_i)\,.
\end{eqnarray*}
The Lie derivative of   the sum $h+g$ and product  $h\cdot g$
functions obey  the familiar rules
\begin{eqnarray}
\lie_F(h+ g) & = & \lie_F(h)+\lie_F(g) \label{eq:liesum}\\
\lie_F(h\cdot g) & = & h\cdot \lie_F(g)+\lie_F(h)\cdot
g\,.\label{eq:lieprod}
\end{eqnarray}
Note that $\lie_F(x_i)=f_i$. Moreover if $p\in \mathbb{R}_d[\vecx]$
then $\lie_F(p)\in
\mathbb{R}_{d+d'}[\vecx]$, %is    a polynomial of degree at most
%$d+d'$,
for some integer $d'\geq 0$ that depends on $d$ and on $F$. This
allows us to view the Lie derivative  of   polynomials along a
polynomial field $F$  as a purely syntactic mechanism, that is as a
function $\lie_F: \field[\xx]\rightarrow \field[\xx]$ that does not
assume anything about the solution of \eqref{eq:ivp}. Informally, we
can view $p$ as a program, and taking the  Lie derivative of $p$ can be
interpreted as  unfolding the definitions of the variables $x_i$'s,
according to the equations in \eqref{eq:ivp} and to the formal rules
for product and sum derivation, \eqref{eq:liesum} and
\eqref{eq:lieprod}. We will pursue this view systematically in
Section \ref{sec:coalg}.

\begin{exa}\label{ex:running0}{%\em
Consider $N=4$, $\xx=(x,y,z,w)^T$ and the set of polynomials $\field[\xx]$.
 The vector field $F=(xz+ z, yw+z,  z,  w)^T$
%(for   nonzero constants $a,b \in \mathbb{R}$)
and the initial condition $v_0=(0,0,1,1)^T$ together define an
initial value problem (with no particular physical meaning)
$\ivp=(F,v_0)$. This problem can be equivalently written in the form
%{ \small
\begin{equation}\label{eq:chem}
\left\{\begin{array}{rcl}
{\dot x(t)} & = & x(t)z(t)+z(t)\\
{\dot y(t)} & = & y(t)w(t)+z(t)\\
{\dot z(t)} & = & z(t)\\
{\dot w(t)} & = & w(t)\\
\vecx(0) & = & v_0\,=\,(0,0,1,1)^T\,.
\end{array}
\right.
\end{equation}
As an example of Lie derivative, if $p=2xy^2+wz$, we have
$\lie_F(p)=4wxy^2 + 2wz + 2xy^2z + 4xyz + 2y^2z$.}
\end{exa}

%The Lie derivative notation is extended to vectors  of functions as expected: if $G=(g_1,....,g_m)$ then $\lie_F(G)=(\lie_F(g_1),...,\lie_F(g_m))$.
%The subscript ${}_F$ will be omitted when clear from context.

The connection between time trajectories, polynomials and Lie
derivatives can be summed up as follows. For  any polynomial $p\in
\field[\vecx]$,  the function $p\circ \vecx(t)  :D\rightarrow
\field$, obtained by composing $p$ as a
function $\reals^N\rightarrow \reals$ with the time trajectory $\vecx(t):D\rightarrow \reals^N$, is    {analytic}: %, that is,   admits a Taylor expansion in a neighborhood of $t=0$.
 we   let $p(\xx(t))$ denote
  the extension of this function %function defined by the Taylor expansion of $p(\vecx(t))$,
 over the largest symmetric open interval of convergence
 (possibly coinciding with $\field$) of its Taylor expansion from 0.
% will refer to
%here we take $U$ to be the
%largest symmetric open interval centered at 0 where the Taylor
%series of $f$  converges (possibly $U=\field$)
We will call $p(\xx(t))$     the \emph{polynomial behaviour induced
by $p$ and by} the initial value problem \eqref{eq:ivp}.  The
connection between Lie derivatives of $p$ along $F$ and the initial
value problem \eqref{eq:ivp} is given by the following equations,
which can be readily checked. Here and in the sequel, we let
$p(v_0)$ denote the real number obtained by evaluating $p$ at $v_0$.
\begin{eqnarray}
p(\xx(t))_{|t=0} & = & p(v_0)\label{eq:initial}\\
\label{eq:lieder} \frac{d\,}{dt}p(\xx(t))& = &
(\lie_F(p))(\xx(t))\,.
\end{eqnarray}
More generally, defining inductively $\lie^{(0)}_F(p)\defin p$ and
$\lie^{(j+1)}_F(p)\defin \lie_F(\lie^{j}_F(p))$, we have the
following equation  for the $j$-th derivative of $p(\xx(t))$
($j=0,1,...$)
\begin{eqnarray}
%p(0) & = & p(v_0)\label{eq:initial}\\
\label{eq:liederj} \frac{d^j\,}{dt^j}p(\xx(t))& = &
(\lie^{(j)}_F(p))(\xx(t))\,.
\end{eqnarray}
In the sequel, we shall often abbreviate $\lie^{(j)}_F(p)$ as
$p^{(j)}$,  and shall omit the subscript ${}_F$ from $\lie_F$ when
clear from the context.

\section{Coalgebraic semantics of  polynomial \ode's}\label{sec:coalg}
In this section we show how to endow the
polynomial ring with a transition  relation structure, hence giving
rise to   coalgebra. Bisimilarity in this coalgebra will correspond
to   equality between polynomial behaviours.

We recall that a  \emph{(Moore) coalgebra}  (see e.g. \cite{Rutten})
with outputs in a set $O$ is a triple $C=(S,\delta,o)$ where $S$ is
a set of \emph{states}, $\delta:S\rightarrow S$ is a
\emph{transition} function, and $o:S\rightarrow O$ is an
\emph{output} function. A \emph{bisimulation} in $C$ is a binary
relation $R\subseteq S\times S$ such that whenever $s\, R\,t$ then:
(a) $o(s)=o(t)$, and (b) $\delta(s)\,R\,\delta(t)$. It is an (easy)
consequence of the general theory of bisimulation that a largest
bisimulation over $S$, called bisimilarity and denoted by $\sim$,
exists, is the union of all bisimulation relations, and     is an
equivalence relation over $S$.

Given an initial value problem $\ivp=(F,v_0)$ of the form
\eqref{eq:ivp}, the triple
\[
C_{\ivp}\defin (\field[\vecx], \lie_F, o)
\]
forms a  {coalgebra} with outputs in $\field$, where: (1)
$\field[\vecx]$  is the set of  {states}; (2) $\lie_F$ acts as the
{transition} function;  and (3) $o$ defined as $o(p)\defin p(v_0)$ is
the   {output} function.  Note that this definition of   coalgebra
is merely syntactic, and does not presuppose anything about the
solution of the given initial value problem. When the standard
definition of bisimulation over coalgebras is instantiated to
$C_{\ivp}$, it yields the following.

\begin{defi}[\lbisim-bisimulation $\sim_{\ivp}$]\label{def:bisim} Let $\ivp$ be  an initial value problem.
 A binary relation   $R\subseteq \field[\vecx]\times \field[\vecx]$ is a \emph{\lbisim-bisimulation} if, whenever $p\,R\,q$ then: (a) $p(v_0)=q(v_0)$, and (b) $\lie(p) \,R\, \lie(q)$. The largest \lbisim-bisimulation over $\field[\vecx]$  is denoted by $\sim_{\ivp}$.
\end{defi}

%It is an (easy) consequence of the general theory of bisimulation that: $\sim$ is the union of all bisimulation relations, and that it is an equivalence relation over $\field[\vecx]$.
We now   introduce   a new coalgebra with outputs in $\field$. Let
$\A$  denote the family of real valued  functions $f$ such  that $f$
is
analytic at 0 and $f$'s domain of definition coincides with % each of
%which we take to be defined over
% $f:U\rightarrow \field$ that
%are analytic in a neighborhood of 0; here we take $U$ to be the
the open interval of convergence  of its Taylor series (nonempty,
centered at 0, possibly coinciding with
$\field$)\footnote{Equivalently, $\A$ is the set of   power series
$f(t)=\sum_{j\geq 0}a_j t^j$ with a positive radius of
convergence.}.
 %the symmetric  open interval  $U$ containing 0 (that is,  each component of $f$ is analytic as a function $U\rightarrow \field$);
We define the coalgebra of analytic functions as
\[
C_{\mathrm{an}}\defin (\A,(\cdot)',o_{\mathrm{an}})
\]
where
%(1) $\mathcal{A}$ is the family of   functions $f:U\rightarrow \field^N$ that are analytic in some open interval $U$ containing 0 (that is,  each component of $f$ is analytic as a function $U\rightarrow \field$); (2)
$(f)'=\frac {df}{dt}$ is the standard  derivative,   and
$o_{\mathrm{fin}}(f)\defin f(0)$ is the output function. We recall
that a \emph{morphism} $\mu$ between two coalgebras with outputs in
the same set, $\mu:C_1\rightarrow C_2$, is a function from states to
states that preserves transitions ($\mu(\delta_1(s))=
\delta_2(\mu(s))$) and outputs ($o_1(s)=o_2(\mu(s))$). It is a standard (and easy) result of coalgebra that a morphism maps bisimilar states into bisimilar states: $s\sim_1s'$ implies $\mu(s)\sim_2\mu(s')$.

%Let us say that a coalgebra with outputs in $\field$ is \emph{convergent} if for each state $s$, the power series $\sum_{j\geq 0} \delta^{(j)}x^j$ has a nonzero radius of convergence.
The coalgebra $C_{\mathrm{an}}$ has a special status, in that, given
any coalgebra  $C$ with outputs in $\field$, \emph{if} there is a
morphism from $C$ to $C_{\mathrm{an}}$, this  morphism  is
guaranteed to be \emph{unique}. For our purposes, it is enough to
focus on $C=C_{\ivp}$. We define the function
$\mu:\field[\xx]\rightarrow \mathcal{A}$ as
\begin{eqnarray*}
\mu(p)& \defin & p(\xx(t))\,.
\end{eqnarray*}
%where  we stipulate that the right hand side  denotes the unique function in $\A$ that extends the polynomial behaviour $p(\xx(t))$.  %Since morphisms map bisimilar states into bisimilar states, this also implies that any two bisimilar states in $C$ are mapped by the final morphism to one and the same state in $C_{fin}$.

\begin{thm}[coinduction]\label{th:finality}
$\mu$ is the unique morphism from $C_{\ivp}$ to $C_{\mathrm{an}}$.   Moreover, the following \emph{coinduction} principle is valid:  $p\sim_\ivp q$ in $C_{\ivp}$   if and only if $p(\xx(t))=q(\xx(t))$ in $\A$. %$p(\xx(t))=q(t)$ in $C_{fin}$.
\end{thm}
\iflncs \else
\begin{proof} %For any $s$ in $C$, set $h(s)=f$, where $f(x)=\sum_{j\geq 0} o(\delta^{(j)})x^j$. By assumption, $f \in \mathcal{A}$. It is readily checked that $h$ is indeed a morphism from $C$ to $C_{fin}$.
The function $\mu$ given above is well defined, because
$p(\xx(t))\in \mathcal{A}$, and is a morphism: for output and
transition preservation, use \eqref{eq:initial} and
\eqref{eq:lieder}, respectively.  By the above recalled standard
result in coalgebra, then $p\sim_\ivp q$ implies $p(\xx(t))\sim
q(\xx(t))$ in $C_{\mathrm{an}}$. Assume now that   $\nu$ is a
morphism from $C_{\ivp}$ to $C_{\mathrm{an}}$. From the definition
of morphism and bisimulation, it is readily checked that for each
$p$, $\mu(p)\sim \nu(p)$ in $C_{\mathrm{an}}$. Finally, we check
that $\sim$ in $C_{\mathrm{an}}$ coincides with equality: indeed, if
two functions
 are bisimilar in $\A$, then they have the same Taylor coefficients (this
is shown by induction on the order of the derivatives, relying on the fact that $f\sim g$ means $f(0)=g(0)$ and
$f'\sim g'$); the
vice-versa is obvious. This completes the proof of both   parts of
the statement.
%that is For the first part, it is enough to show that $\sim$ in $C_{fin}$ is the identity. This follows from the fact that two analytic functions in $\mathcal{A}_U$ are bisimilar in $C_{fin}$ if and only if they have the same Taylor expansion at 0; but this in turn is the case if and only if they coincide as functions on $U$.
\end{proof}
\fi

\begin{rem}[categorical presentation]\label{rem:cat}{%\em
Existence of a morphism from any coalgebra  with outputs in $\reals$
into $C_{\mathrm{an}}$  is not guaranteed:  for instance, the
  sequence  (\emph{stream})  of coefficients of any series with a radius of
convergence 0, such as $(0!,1!,2!,...,i!,...)$,  trivially induces a
coalgebra from which no morphism to $C_{\mathrm{an}}$ exists. In
this sense, $C_{\mathrm{an}}$ is not \emph{final}. Note that
$C_{\mathrm{an}}$ can be injected into the coalgebra of streams in
the sense of Rutten \cite{Rutten}, which is indeed final.

In a more categorical perspective,   $C_{\mathrm{an}}$ induces a
so-called \emph{covariety} -- see e.g. \cite{covariety} -- that is,
the category of coalgebras that have a unique morphism into
$C_{\mathrm{an}}$:  $C_{\mathrm{an}}$ is of course final in this
covariety. How to
  characterise such covariety more   explicitly, for
example    in terms of comonads, is left for future research. }
\end{rem}

%\begin{prop}\label{prop:morph} The function $h:C_{pol}\rightarrow C_{fin}$ defined as $h(p)=p(\xx(t))$ for each $p\in \field[\vecx]$ is the unique morphism from $C_{pol}$ to $C_{fin}$.
%\end{prop}
%\begin{proof} The function $h$ given in the statement is well defined (because $p(\xx(t))\in \mathcal{A}$) and is a morphism (for transition preservation, use \eqref{eq:lieder}); hence it is unique by virtue of Theorem \ref{th:finality}.
%\end{proof}

Theorem \ref{th:finality}  allows one to prove polynomial relations
among the components $x_i(t)$ of $\vecx(t)$, say that
$p(\xx(t))=q(\xx(t))$, by {coinduction}, that is, by exhibiting a
suitable \lbisim-bisimulation relating the polynomials $p$ and $q$.

\begin{exa}\label{ex:sin}{%\em
For $N=2$, consider the vector field  $F=(x_2,-x_1)^T$  with the initial value $v_0=(0,1)^T$.
 The binary relation $R\subseteq \field[x_1,x_2]\times \field[x_1,x_2]$ defined thus
\[
R = \{ \,(0,0),\,(x_1^2+x_2^2,1)\,\}
\]
is easily checked to be an \lbisim-bisimulation. Thus we have proved
the polynomial relation  $x_1^2(t)+x_2^2(t)=1$. Note that the unique
solution to the given  initial value problem is the pair of
functions $\xx(t)=(\sin(t),\cos(t))^T$. This way we have proven the
familiar trigonometric identity $\sin(t)^2+\cos(t)^2=1$.}
\end{exa}

This proof method can be greatly enhanced by a so called
\emph{\lbisim-bisimulation up to} technique, in the spirit of
\cite{San}. \iflncs See \cite{Full}. \else This can be regarded as a
form of up-to context technique, since we are just considering the
closure w.r.t. the syntax, that is, the algebraic structure of the
polynomial ring.

\begin{defi}[\lbisim-bisimulation up to]\label{def:bisimupto} Let    $R\subseteq \field[\vecx]\times \field[\vecx]$ be a binary relation.
Consider the binary relation $\widehat R$ defined by:
$p \,\widehat R\, q$   iff    there are $m\geq 0$ and    polynomials
$h_i,p_i,q_i$ ($i=1,...,m$) such that: $p=\sum_{i=1}^m h_i p_i$ and
$q=\sum_{i=1}^m h_i q_i$ and   $p_i \,R\, q_i$, for $i=1,...,m$.

A relation $R\subseteq \field[\vecx]\times \field[\vecx]$ is a
\emph{\lbisim-bisimulation up to} if, whenever $p\,R\,q$ then: (a)
$p(v_0)=q(v_0)$, and (b)    $\lie(p)  \,\widehat R\, \lie(q)$.
\end{defi}

Note that, from the definition, for each relation $R$ we have
$R\subseteq \widehat R$.

\begin{lem} Let $R$ be an \lbisim-bisimulation up to. Then $\widehat R$ is an \lbisim-bisimulation,
 consequently  $  R\,\subseteq\,\sim_{\ivp}$.
\end{lem}
\iflncs\else
\begin{proof} In order to check that $\widehat R$ is an \lbisim-bisimulation, assume  that $p \,\widehat R\,q$, that is  $p=\sum_{i=1}^m h_i p_i$ and  $q=\sum_{i=1}^m h_i q_i$, for some $h_i,p_i,q_i$ ($i=1,...,m$) as specified by Definition \ref{def:bisimupto}.  It is immediate to check that $p(v_0)=q(v_0)$, which proves condition (a) of the definition of \lbisim-bisimulation.  Furthermore, for each $i=1,..,m$, we have by assumption that $\lie(p_i)\, \widehat R \,\lie(q_i)$. That is, for suitable $g_{ij}$'s and $r_{ij}$'s, we have
\begin{align*}
\lie(p_i) & =   \sum_j g_{ij} r_{ij} & \lie(q_i) & =  \sum_j g_{ij} s_{ij}&  r_{ij}\, & R\,  s_{ij}\,.
\end{align*}
Recalling the rules  for the Lie derivative, \eqref{eq:liesum} and
\eqref{eq:lieprod}, we   have {\small
\begin{align*}
\lie(p) & = \; \sum_{i}  h_i \lie(p_i)+\lie(h_i)p_i & & =  \sum_{i}  h_i \sum_j g_{ij} r_{ij}+\lie(h_i)p_i &  =  \sum_{i}\sum_{j} h_i g_{ij} r_{ij} +\sum_{i} \lie(h_i)p_i\\
        & \widehat R  \; \sum_{i}\sum_{j} h_i g_{ij} s_{ij} +\sum_{i} \lie(h_i)q_i & & =   \sum_{i} h_i\sum_{j} g_{ij} s_{ij} +\sum_{i} \lie(h_i)q_i  & =  \sum_{i}  h_i \lie(q_i)+\lie(h_i)q_i\hspace*{1.1cm}\\
        & =  \; \lie(q)\hspace*{\fill}\mbox{\ }\,.
\end{align*}}

\noindent
This proves condition (b) of the definition of
\lbisim-bisimulation.
\end{proof}
\fi

\begin{exa}\label{ex:running1}{%\em
%Consider $N=4$, the set of polynomials $\field[x,y,z,w]$,  the vector field $F=(a xz+b z,a xz+w, b z, a w)$ (for   nonzero constants $a,b \in \mathbb{R}$) and an initial condition $v_0=(0,0,1,1)^T$. This defines an initial value problem related to a certain chemical reaction.
Consider the initial value problem of Example \ref{ex:running0}
%. Let
%$Id$ denote the binary identity relation on $\field[\xx]$ and
%consider
and the   relation defined below
\[
R\;= \; %Id\;\cup\;
\{\,(xz^j,yw^j)\,,\,(z^j,w^j)\,:\,j\geq
0\}\,.
\]
It is easy to check that $R$ is an \lbisim-bisimulation up to.
\iflncs\else
 As an
example, let us check condition (b) for a pair     $(xz^j,yw^j)$:
{\small
\[
\lie(xz^j)\,=\,(xz + z)z^j+jxz^j\,=\,
(xz^{j+1}+jxz^j+zz^{j})\;\widehat R\; (yw^{j+1}+jyw^j+zw^{j})\,
=\,(yw + z)w^j+jyw^j\,=\,\lie(yw^j)\,.
\]
}
\noindent
This proves that $x(t)=y(t)$ and that $z(t)=w(t)$.
\fi
}\end{exa}
In the next two sections we will prove that this technique can be  fully automated by resorting to the concept of \emph{ideal} in a polynomial ring. %For now, we give a more general version of bisimulation up to. Let us fix a
\fi

\section{Algebraic characterization of \lbisim-bisimilarity}\label{sec:bisim}
We first review the notion of polynomial ideal from Algebraic
Geometry, referring the reader to e.g. \cite{Cox} for a
comprehensive treatment. A set of polynomials $I\subseteq
\field[\xx]$ is an \emph{ideal} if: (1) $0\in I$, (2) $I$ is closed
under sum +, (3) $I$ is absorbing under product $\cdot$, that is
$p\in I$ implies $h\cdot p\in I$ for each $h\in \field[\xx]$. Given
a set of polynomials $S$, the ideal generated by $S$, denoted by
$\ide S$, is defined as {\small
\begin{equation}\label{eq:ide}
%\ide S & \defin &
\left\{\sum_{j=1}^m   h_jp_j\,: \, \text{    $m\geq
0$, $h_j\in\field[\xx]$ and $p_j\in S$, for }j=1,...,m \right\}\,.
\end{equation}}
The polynomial coefficients $h_j$ in  the above definition are
called \emph{multipliers}.  It is clear that $\ide S$ is the
smallest ideal containing $S$, which implies that $\ide{\ide S}
=\ide S$. Any set $S$ such that $\ide S=I$ is called a \emph{basis}
of  $I$. Every ideal in the polynomial ring $\field[\xx]$ is finitely generated,
that is has a finite basis (a version of Hilbert's basis theorem). %The key to
%connecting binary relations and    ideals is given by the following
%definition and lemma.

%A binary relation is completely characterized by its kernel, in the
%following sense.
%\begin{lem}\label{lemma:kernel} For any binary relation $R$, $p\,\widehat R\,q$ if and  only if $p-q\in \ide{\ker(R)}$.
%\end{lem}

\lbisim-bisimulations can be connected to certain types of ideals.
This connection relies on  Lie  derivatives. First, we define the
Lie derivative of any set $S\subseteq \field[\xx]$   as follows
\begin{eqnarray*}
\lie(S) & \defin & \{\lie(p): p\in S\}\,.
\end{eqnarray*}
We say that   $S$ is a \emph{pre-fixpoint} of $\lie$ if
$\lie(S)\subseteq S$. \lbisim-bisimulations  can be characterized as
particular pre-fixpoints of $\lie$  that are also ideals,
called \emph{invariants}.
%In the sequel, we will consider the
%following set of
%polynomials\footnote{This definition depends on the $v_0$ in the given initial value problem \eqref{eq:ivp},
%which however is left out of the notation.}
%\begin{eqnarray*}
%\Izero   & \defin &   \{p:\,p(v_0)=0\}\,.
%\end{eqnarray*}

\begin{defi}[invariant ideals]\label{def:invariant}
%\begin{eqnarray*}
%X_0 & \defin & \{p: p(\xx(t))=0\text{ for each }t\in U\}\,.
%\end{eqnarray*}
Let $\ivp=(F,v_0)$. An ideal $I$ is a $\ivp$-\emph{invariant} if:
(a) $p(v_0)=0$ for each $p\in I$, and (b) $I$ is a pre-fixpoint of
$\lie_F$.
\end{defi}

We will drop the $\ivp$- from \emph{$\ivp$-invariant}   whenever
this is
clear from the context. The following definition and lemma provide the link between invariants and \lbisim-bisimulation.
%In the following lemma, the \emph{only if}
%part also holds when weakening \emph{bisimulation} to\emph{
%bisimulation up to}.

\begin{defi}[kernel] The \emph{kernel} of a binary relation $R\subseteq \reals[\xx]\times \reals[\xx]$ is $\ker(R)\defin \{p-q: p\,R\,q\}$.
\end{defi}

\begin{lem}\label{lemma:prefixp} Let $R$ be a binary relation.
If $R$ is an \lbisim-bisimulation then
$\ide{\ker(R)}$ is an invariant.
Conversely, given an invariant $I$, then $R=\{(p,q):p-q\in I\}$ is an \lbisim-bisimulation.
\end{lem}

%Thus the task of checking $p\sim_{F} q$ can be reduced to building an invariant that contains $p-q$.
Consequently, proving  that  $p\sim_{\ivp} q$ is equivalent to
exhibiting an invariant $I$ such  that $p-q\in I$.
\iflncs\else

\begin{exa}\label{ex:running2}{%\em
Consider the initial value
problem of Example \ref{ex:running1}. Let $I=\ide{\{x-y,z-w\}}$. Let us check  that $I$ is an invariant.
  Let $p=h_1(x-y)+h_2(z-w)$ be a generic element of $I$. Clearly $p(v_0)=0$, thus condition (a) is satisfied.
    Concerning (b), we consider the two summands separately:
    {\small
\begin{eqnarray*}
\lie(h_1(x-y)) &= & \lie(h_1)(x-y)+h_1\lie(x-y)\\
               & = & \lie(h_1)(x-y)+h_1(xz+ z-(yz+ w))\\
               & = & \big(\lie(h_1)+h_1 z\big)(x-y)+ h_1(z-w)\\
               & \in & I\\
\lie(h_2(z-w)) &= & \lie(h_1)(z-w)+h_1\lie(z-w)\\
               & = & \lie(h_1)(z-w)+h_1(z-w)\\
               & = & \big(\lie(h_1)+h_1\big) (z-w)\\
               &\in & I\,.
\end{eqnarray*}
}
Consequently, $\lie(p)=\lie(h_1(x-y))+\lie(h_2(z-w))\in I$.
}
\end{exa}

\fi
%It is a general result of ideals theory that any ideal  has a \emph{finite} set of generators $B$, that is a set $B$  such that $\ide B = I$: this is Hilbert's theorem, see e.g. \cite{Cox}.
A more general problem than equivalence checking  is   finding
\emph{all} valid  polynomial equations of a given form. We will
illustrate an algorithm to this purpose in the next section.
%decomposed into: (1) finding a finite set of generators $B$ of a suitable invariant containing; (2) check  that $p-q$ can be generated from $B$. The latter task is known as the \emph{ideal membership} problem. It can be solved  effectively by turning  $B$ into  a special form,  called \emph{Gr\"{o}bner} basis. There are effective procedures to do this, such as Buchberger's algorithm; we refer the reader to \cite{Cox} for details. Overall, these ingredients lead to a decision scheme, that we will illustrate in the next section.

The following result sums up the different characterization of
\lbisim-bisimilarity  $\sim_{\ivp}$.
 %in a few different,   equivalent, ways.
%This characterization may be instructive, although the following result will not be invoked in the rest of the paper.
In what follows, we will   denote the constant zero function in $\A$
simply by $0$ and    consider the following set of
 polynomials.
 \begin{eqnarray*}
%I_0 & \defin & \{p:\,p(v_0)=0\}\,.
\Zzero & \defin & \{p: p(\xx(t))\mbox{ is identically }0\,\}\,.
\end{eqnarray*}
%Note that $\Zzero\subseteq \Izero$.
The following result also proves that $\Zzero$ is the largest
$\ivp$-invariant.
%We also denote the constant zero function in $\A$ simply by $0(t)$.

\iflncs
\begin{thm}[\lbisim-bisimilarity via ideals]\label{th:bisim} %We have
%the following characterizations of   \lbisim-bisimilarity.
For any pair of polynomials $p$ and $q$:
$p\,\sim_{\ivp}\,q  \; \text{ iff }\; p-q  \in \ker(\sim_{\ivp})
                    =    \Zzero\label{eq:c1} $
                  $ \overset{(1)}{=}   \left\{p: \,p^{(j)}(v_0)=0 \text{ for each }j\geq 0 \right\}
                   =   \bigcup\left\{I:\,I \text{ is a $\ivp$-invariant } \right\}$.
%\\ & = & \mu^{-1}\left(0\right)\label{eq:c4}
\end{thm}
\else
\begin{thm}[\lbisim-bisimilarity via ideals]\label{th:bisim} We have
the following characterizations of   \lbisim-bisimilarity. For any pair of polynomials $p$ and $q$:
{\small
\begin{eqnarray}
p\,\sim_{\ivp}\,q  \; \text{ iff }\; p-q& \in& \ker(\sim_{\ivp})\label{eq:c0} \\
                    & = &  \Zzero\label{eq:c1}\\
                 & = &  \left\{p: \,p^{(j)}(v_0)=0 \text{ for each }j\geq 0 \right\}
                 \label{eq:c3}\\
                 & = & \bigcup\left\{I:\,I \text{ is a $\ivp$-invariant }
                 \right\}\label{eq:c2}\,.
%\\ & = & \mu^{-1}\left(0\right)\label{eq:c4}
\end{eqnarray}}
\end{thm}
\fi

% It is easily checked that $\lie(I)$ is in turn an ideal.

\section{Computing invariants}\label{sec:algorithm}
By Theorem \ref{th:bisim}, proving $p\sim_\ivp q$ means finding an
invariant $I$ such that $p-q\in I\subseteq \Zzero$. More generally, we
focus here on the problem of finding invariants that include a user-specified set of   polynomials.
In the sequel, we will make use of the
following  two basic  facts about ideals, for whose proof we refer
the reader to \cite{Cox}.
\begin{enumerate}
%\item
\item Any infinite ascending chain of ideals in a polynomial ring, $I_0\subseteq I_1\subseteq \cdots$,
stabilizes at some finite $k$. That is, there is $k\geq 0$ such that
$I_k=I_{k+j}$ for each $j\geq 0$. This is just another version of
Hilbert's basis theorem.
\item The \emph{ideal membership problem}, that is, deciding whether $p\in I$, given $p$  and a finite
set of $S$ of \emph{generators}  (such that $I=\ide S$), is decidable (provided the
coefficients used in $p$ and in $S$ can be finitely represented), although it requires exponential space in the number of variables.
 The ideal membership will be  further discussed later on in the section.
\end{enumerate}
The main idea is introduced by the   naive algorithm presented
below.

\iflncs
\subsubsection{A naive algorithm}
\else
\subsection{A naive algorithm}
\fi Suppose we want to decide whether  $p\in \Zzero$. It is quite
easy to devise an  algorithm that  computes the smallest invariant
containing $p$, or returns `no' in case no such invariant exists,
i.e. in case $p\notin \Zzero$. Consider the successive Lie
derivatives of $p$, $p^{(j)}= \lie^{(j)}(p)$ for $j=0,1,...$. For
each $j\geq 0$, let $I_j\defin \ide{\{p^{(0)},...,p^{(j)}\}}$. Let
$m$ be the least integer such that either \iflncs \vspm
{\small\center (a) $p^{(m)}(v_0)\neq 0$, \ \ \ \ \ or\ \ \ \ \ (b)
$I_m=I_{m+1}$.

\noindent
}
\else
\begin{itemize}
\item[(a)] $p^{(m)}(v_0)\neq 0$, or
\item[(b)] $I_m=I_{m+1}$.
\end{itemize}
\fi
If (a) occurs, then $p\notin \Zzero$, so we return `no'
 (Theorem \ref{th:bisim}\iflncs(1)\else\eqref{eq:c3}\fi); if (b) occurs, then $I_m$ is the least invariant containing $p$.
 Note that the integer $m$ is well defined:  $I_0\subseteq I_1\subseteq I_2\subseteq \cdots$ forms
  an infinite ascending chain of ideals, which must stabilize in a finite number of steps
  (fact 1 at the beginning of the section).
In particular, as soon as $I_{i+1}=I_i$ the chain gets stable, as a
consequence of the derivation rules    \eqref{eq:liesum} and
\eqref{eq:lieprod}.
  %, and it is a general result of ideal theory that any such chain in a polynomial ring must stabilize at some finite $m$ (see \cite{Cox}).

Checking condition (b) amounts to deciding if $p^{(m+1)}\in I_m$.
This is an instance of the  {ideal membership} problem, which can be
solved effectively.  Generally speaking, given a polynomial $p$ and
finite set of polynomials $S$,  deciding the ideal membership $p\in
I=\ide S$ can be accomplished by  first   transforming   $S$ into a
\emph{Gr\"{o}bner basis}   $G$ for $I$
 (via, e.g. the Buchberger's algorithm),    then computing $r$,
  the \emph{residual} of $p$ modulo $G$  (via a sort
generalised division of $p$ by $G$): one has that $p\in I$  if and
only if $r=0$ (again, this procedure can be carried out effectively
only if the coefficients involved in $p$ and $S$ are finitely
representable; in practice, one often confines to rational
coefficients). We refer the reader to \cite{Cox} for further details
on the ideal membership problem. Known procedures to compute
Gr\"{o}bner bases have exponential worst-case space complexity depending on the number of variables,  although    may perform
reasonably well in some concrete cases. One
should in any case invoke such   procedures parsimoniously.
%that can be solved by first  computing a Gr\"{o}bner basis $G$ of $I_j$ from $\{p_0,...,p_j\}$, which can be done by any of the known algorithms, and then checking if $p_{j+1}\bmod G =0$, for which again algorithms are available.
\iflncs\else
Here is a small example to illustrate the above outlined algorithm.

\begin{exa}\label{ex:running3}{%\em
Consider again the initial value problem  of Example \ref{ex:running1}.
Let $p=x-y$. With the help of a computer algebra system, we can easily check the following.
{%\small
\begin{itemize}
\item $p^{(0)}=p$ and $p^{(0)}(v_0)=0$;
\item $p^{(1)}= xz - yw $, $p^{(1)}(v_0)=0$ and $p^{(1)}\notin I_0=\ide{\{p^{(0)}\}}$;
\item $p^{(2)}=-w^2 y - w y - w z + x z^2 + x z + z^2$, $p^{(2)}(v_0)=0$ and $p^{(2)}\notin I_1=\ide{\{p^{(0)},p^{(1)}\}}$
\item $p^{(3)}=-w^3 y - 3 w^2 y - w^2 z - w y - 3 w z + x z^3 + 3 x z^2 + x z + z^3 + 3 z^2$, $p^{(3)}(v_0)=0$ and finally\footnote{Indeed, a Gr\"{o}bner
basis for $I_2$ is $G=\{x - y, yz -wy  , z^2-w z    \}$, and $p^{(3)}=(z^3 + 3 z^2 + z)(x-y)+(w^2 + w z + 3 w + z^2 + 3 z + 1)(yz-wy)
+ (w + z + 3)(z^2-w z)$.} $p^{(3)}\in
I_2=\ide{\{p^{(0)},p^{(1)},p^{(2)}\}}$.
\end{itemize}
}
\noindent
Hence $I_2$
is the least invariant containing $p=x-y$, thus proving that $x-y\in \Zzero$.
}
\end{exa}

\fi
We will introduce below a more general algorithm, which can also deal with   (infinite) \emph{sets} of user-specified polynomials.
First, we
need to introduce the concept of template.

\iflncs
\subsubsection{Templates}\label{sub:templ}
\else
\subsection{Templates}\label{sub:templ}
\fi
Polynomial templates have been introduced by   Sankaranarayanan, Sipma and Manna  in \cite{San04} as a means to
compactly specify   sets of polynomials.  Fix a tuple of $n\geq 1$ of
distinct \emph{parameters}, say $\aa=(a_1,...,a_n)$, disjoint from
$\xx$. Let $Lin(\aa)$, ranged over by $\ell$, be the set of
\emph{linear expressions} with coefficients in $\field$ and
variables in $\aa$; e.g. $\ell=5a_1 +42a_2-3a_3$   is one such
expression\footnote{Differently from Sankaranarayanan et al. we do
not allow linear expressions with a constant term, such as $2+5a_1
+42a_2-3a_3$. This  minor syntactic restriction does not practically
affect the expressiveness of the resulting polynomial templates.}. A
\emph{template} is a polynomial in $Lin(\aa)[\xx]$, that is, a
polynomial with linear expressions as coefficients; we let $\pi$
range over templates. For example, the following is a template:
\iflncs $ \pi  =   (5a_1+(3/4) a_3)xy^2+ (7a_1+(1/5) a_2)xz + (a_2+42a_3)$.
\else
{
\begin{eqnarray*}
\pi & = & (5a_1+(3/4) a_3)xy^2+ (7a_1+(1/5) a_2)xz + (a_2+42a_3)\,.
\end{eqnarray*}}
\fi
Given a  vector $v=(\lambda_1,...,\lambda_n)^T\in \field^n$, we will let
$\ell[v]\in \field$ denote the result of replacing each parameter
  $a_i$ with $\lambda_i$, and evaluating the resulting expression;
we will let $\pi[v]\in \field[\xx]$ denote the polynomial obtained
by replacing each $\ell$ with $\ell[v]$ in $\pi$.  Given a set
$S\subseteq \field^n$, we let $\pi[S]$ denote the set
$\{\pi[v]\,:\,v\in S\}\subseteq \field[\xx]$. %Notationally, when no
%confusion arises, we will sometimes  write $\pi[\field^n]$ simply as
%$\pi$.

The (formal) Lie derivative of $\pi$ is defined as expected, once
linear expressions are treated as constants; note that $\lie(\pi)$
is still a template.
% if $\pi=\sum_{\alpha\in M} \ell_\alpha \alpha$, for some multiset $M$ of monomials, then $\lie(\pi)\defin \sum_{\alpha\in M} \ell_\alpha\lie(\alpha)$, which is still a template.
 It is easy to see that the following property is true: for  each $\pi$ and $v$, one has $\lie(\pi[v])   =   \lie(\pi)[v]$. This property extends as expected to the $j$-th Lie derivative ($j\geq 0$)
\begin{eqnarray}
\lie^{(j)}(\pi[v]) & = & \lie^{(j)}(\pi)[v]\,.\label{eq:templder}
\end{eqnarray}
%This property .

\iflncs
\subsubsection{A double chain algorithm}\label{sub:algo}
\else
\subsection{A double chain algorithm}\label{sub:algo}
\fi
%We present an algorithm that computes a most general template  all
%of whom instances are included in $\Zzero$, and the most economical
%invariant witnessing this inclusion.
%More precisely,
We present an algorithm that, given a template $\pi$ with $n$
parameters,    returns a pair $(V,J)$, where $V\subseteq \field^n$
is such that $\pi[\field^n]\cap \Zzero = \pi[V]$, and $J$ is the
smallest invariant  that includes $\pi[V]$. %, possibly $J=\{0\}$.
We first give a purely mathematical description of the algorithm,
postponing its effective representation   to the next subsection.
The algorithm is based on building two chains of sets, a descending
chain of vector spaces and an (eventually) ascending chain of
ideals. The   ideal chain is used to detect the stabilization of the
sequence.  In fact, in the sequence of vector spaces below,
$V_{i+1}=V_i$ does not in itself  imply that $V_{i+k}=V_i$ for each
$k\geq 1$. Formally, for each $i\geq 0$, consider the sets
{\iflncs\small\else\fi
\begin{eqnarray}
V_i & \defin & \{v\in \field^n\,:\,\pi^{(j)}[v](v_0)=0\text{ for }j=0,...,i\,\}\label{eq:Vi}\\
J_i & \defin & \ide{\bigcup_{j=0}^i \pi^{(j)}[V_i]}\label{eq:Ji}\,.
\end{eqnarray}
}

\noindent
It is easy to check that each $V_i\subseteq \field^n$ is a vector
space over $\field$ of dimension $\leq n$. Now let $m\geq 0$ be  the
least integer such that the following conditions are \emph{both}
true:
{\iflncs\small\else\fi
\begin{eqnarray}
V_{m+1} & = &V_m \label{eq:Vm}\\
J_{m+1} & = & J_m\,.\label{eq:Jm}
\end{eqnarray}
}

\noindent
The algorithm returns $(V_m,J_m)$. Note that the integer  $m$ is
well defined: indeed, $V_0\supseteq V_1\supseteq \cdots$ forms  an
infinite descending chain of finite-dimensional vector spaces, which
must   stabilize in finitely many steps. In other words,  we can
consider the least $m'$ such that $V_{m'}=V_{m'+k}$ for each $k\geq
1$.
%Suppose that for no  $j\leq m'$  condition (2) is met  (otherwise $m$ is the least such $j$, by definition).   If $\dim(V_{m'})=0$, then by definition $m=m'$; otherwise,
Then $J_{m'}\subseteq J_{m'+1}\subseteq\cdots$ forms    an infinite
ascending chain of ideals, which must
  stabilize at some $m\geq m'$. % (see the beginning of Section \ref{sec:bisim}).
Therefore there must be some index $m$ such that \eqref{eq:Vm}  and
\eqref{eq:Jm} are both satisfied, and we   choose the least such
$m$.

The next theorem states the correctness and relative  completeness
of this abstract algorithm. Informally, the algorithm will output
the largest space $V_m$ such that $\pi[V_m]\subseteq \Zzero$ and
 the smallest invariant $J_m$ witnessing this inclusion. Note that, while
  typically the user will be interested in
 $\pi[V_m]$,   $J_m$ as well may contain useful information, such
 as higher order, nonlinear conservation laws.
 %For part (b), note that the intersection of an arbitrary collection of invariants  is an invariants
We need a technical lemma.

\begin{lem}\label{lemma:stab}
Let $V_m,J_m$ be the sets returned by the algorithm. Then %sequence  of sets $V_i,J_i$ stabilizes at $m$:  that is,
for each $j\geq 1$, one has
 $V_m=V_{m+j}$ and $J_m=J_{m+j}$.
\end{lem}

\begin{thm}[correctness and relative completeness]\label{th:corr} Let $V_m,J_m$ be the
 sets returned by the algorithm for a polynomial template $\pi$. %We have the following:
\begin{itemize}[align=left]
\item[(a)] $\pi[V_m]=\Zzero\cap \pi[\field^n]$;
\item[(b)] $J_m$ is  the smallest invariant containing $\pi[V_m]$.
\end{itemize}
\end{thm}
\begin{proof}
Concerning part (a), we first note that $\pi[v]\in \Zzero\cap
\pi[\field^n]$ means   $(\pi[v])^{(j)}(v_0)= \pi^{(j)}[v](v_0)=0$
for each $j\geq 0$ (Theorem
\ref{th:bisim}\iflncs(1)\else\eqref{eq:c3}\fi), which, by
definition,
  implies      $v\in V_j$ for each $j\geq 0$, hence
$v\in V_m$. Conversely, if $v\in V_m=V_{m+1}=V_{m+2}=\cdots$ (here
we are using Lemma \ref{lemma:stab}), then by definition
$(\pi[v])^{(j)}(v_0)= \pi^{(j)}[v](v_0)=0$ for each $j\geq 0$, which
implies that $\pi[v]\in \Zzero$  (again Theorem
\ref{th:bisim}\iflncs(1)\else\eqref{eq:c3}\fi). Note that in proving
both inclusions we have  used property \eqref{eq:templder}.

Concerning part (b),  it is enough to prove that: (1) $J_m$ is an
invariant,   (2) $J_m\supseteq \Zzero\cap \pi[\field^n]$,  and (3)
for any invariant $I$ such that $\Zzero\cap \pi[\field^n] \subseteq
I$, we have that $J_m\subseteq I$. We first prove (1),  that  $J_m$
is an invariant. Indeed,  for each $v\in V_m$ and each
$j=0,...,m-1$, we have $\lie(\pi^{(j)}[v])=\pi^{(j+1)}[v]\in J_m$ by
definition, while for $j=m$, since $v\in V_m=V_{m+1}$, we have
$\lie(\pi^{(m)}[v])=\pi^{(m+1)}[v]\in J_{m+1}=J_m$ (note that in
both cases we have used property \eqref{eq:templder}). Concerning
(2),  note that $J_m\supseteq \pi[V_m]=\Zzero\cap \pi[\field^n]$ by
virtue of part (a). %Indeed, for each $v\in \field^n$ such that $\pi[v]\in \Zzero$, we have by definition of $\Zzero$ that $(\pi[v])^{(i)}(v_0)= (\pi^{(i)}[v])(v_0)=0$ for $i=0,1,...$ (again, we have used here property \eqref{eq:templder}). Hence by definition of $V_m$, we have $v\in V_m$, thus $\pi[v]\in J_m$.
Concerning (3),  consider any invariant $I\supseteq \Zzero\cap
\pi[\field^n]$.  We show by induction on $j=0,1,...$ that for each
$v\in V_m$,  $\pi^{(j)}[v]\in I$; this will imply the wanted
statement.  Indeed, $\pi^{(0)}[v]=\pi[v]\in \Zzero\cap
\pi[\field^n]$, as $\pi[V_m]\subseteq \Zzero$ by (a). Assuming now
that $\pi^{(j)}[v]\in I$, by invariance of $I$ we have
$\pi^{(j+1)}[v]= \lie(\pi^{(j)}[v])\in I$ (again, we have used here
property \eqref{eq:templder}).
\end{proof}

According to Theorem \ref{th:corr}(a), given a template $\pi$ and $v\in \field^n$, checking if $\pi[v]\in \pi[V_m]$ is
equivalent to  checking if $v\in V_m$, which can be effectively done
 knowing  a basis $B_m$ of   $V_m$. We show how to effectively compute such a basis in the following.

\iflncs
\subsubsection{Effective representation}\label{sub:eff}
\else
\subsection{Effective representation}\label{sub:eff}
\fi
For   $i=0,1,...$, we have to give effective ways to:
\iflncs (i) represent the sets $V_i, J_i$ in \eqref{eq:Vi} and \eqref{eq:Ji};  and, (ii) check  the termination conditions \eqref{eq:Vm} and \eqref{eq:Jm}.
\else
\begin{enumerate}
\item[(i)] represent the sets $V_i, J_i$ in \eqref{eq:Vi} and \eqref{eq:Ji};  and
\item[(ii)] check  the termination conditions \eqref{eq:Vm} and \eqref{eq:Jm}.
\end{enumerate}
\fi
It is quite easy to address  (i) and (ii)  in the case of the vector
spaces $V_i$.
%By representing $V_i$ as a set of linear constraints over the tuples $v\in \field^n$.
For each $i$, consider the linear expression $\pi^{(i)}(v_0)$.  By
factoring out the parameters $a_1,...,a_n$ in this expression, we
can write, for a suitable (row) vector of coefficients
$t_i=(t_{i1},...,t_{in})\in \field^{1\times n}$: \iflncs
$\pi^{(i)}(v_0)   =   t_{i1}\cdot a_1+\cdots +t_{in} \cdot a_n$
\else  {
\begin{eqnarray*}
\pi^{(i)}(v_0) & = & t_{i1}\cdot a_1+\cdots +t_{in} \cdot a_n %\sum_{j=1}^m t_{ij} a_j\,.
\end{eqnarray*}
}
\fi
%\write $\pi^{(i)}$ as follows,  for some set\footnote{Considering a multiset would give rise to an equivalent constraint.} of monomials $M_i$
%\[
%\pi^{(i)}= \sum_{\alpha\in M_i}\ell_{\alpha,i}\alpha\,.
%\]
The condition on $v\in \field^n$, $\pi^{(i)}[v](v_0)=0$, can then be
translated into the linear constraint on $v$
\vspm{\iflncs\small\else\fi
\begin{eqnarray}\label{eq:constr}
t_i\cdot v & = & 0\,.%\sum_{\alpha\in M_i}\ell_{\alpha,i}\alpha(v_0)=0\,.
\end{eqnarray}}

\noindent
%\vspm
Letting $T_i\in \field^{i\times n}$ denote the matrix obtained  by
stacking the rows $t_1,...,t_i$ on above the other, we see that
$V_i$ is the right null space of $T_i$. That is (here,
$\mathbf{0}_i$ denotes the null vector in $\field^{i}$):
\iflncs $ V_i   =   \{v\in \field^n\,:\, T_i v=\mathbf{0}_i\,\}$. \else
\begin{eqnarray*}
V_i & = & \{v\in \field^n\,:\, T_i v=\mathbf{0}_i\,\}\,.
\end{eqnarray*}
\fi
%Hence $V_i=\{v\in \field^n: \gamma_j[v]=0 \text{ for }j=0,...,i\}$.
Checking whether $V_i=V_{i+1}$ or not     amounts then  to checking
whether  the vector $t_{i+1}$ is or not linearly dependent from the
rows in $T_i$, which can be accomplished by   standard and efficient
linear algebraic techniques. In practice,  the linear constraints
\eqref{eq:constr}    can be resolved and propagated
incrementally\footnote{E.g., if   $\pi=a_1x+a_2y+a_3x+a_4w$ and
$v_0=(0,0,1,1)^T$, then $\pi[v](v_0)=0$ is resolved and propagated
applying to $\pi$ the substitution $[a_3\mapsto -a_4]$.}, as   they
are generated, following the computation of the   derivatives
$\pi^{(i)}$.
Concerning the representation of the ideals $J_i$, we will use the
following lemma\footnote{The restriction that linear expressions in templates do not contain constant terms
is crucial here.}.

\begin{lem}\label{lemma:basisideal}
Let $V\subseteq \field^n$ be a vector space with $B$ as a basis,
and $\pi_1,...,\pi_k$ be templates.  Then %$\ide{\pi[V]}=\ide{\pi[B]}$. Consequently,
$\ide{\cup_{j=0}^k \pi_j[V]}=\ide{\cup_{j=0}^k \pi_j[B]}$.
\end{lem}

Now let $B_i$ be a finite basis of $V_i$, which  can be easily built
 from the matrix $T_i$. By the previous lemma, $\cup_{j=1}^i
\pi^{(j)}[B_i]$ is a \emph{finite} set of generators for $J_i$: this
solves the representation problem.  Concerning the termination
condition, we note that, after checking that actually $V_i=V_{i+1}$,
checking   $J_i=J_{i+1}$ reduces to checking that
\iflncs $\pi^{(i+1)}[B_i] \subseteq   \ide{\cup_{j=1}^i \pi^{(j)}[B_i]}
\;=\;J_i\,(\mathbf{*})$.
\else
\begin{eqnarray}
\pi^{(i+1)}[B_i]&\subseteq & \ide{\cup_{j=0}^i \pi^{(j)}[B_i]}
\;=\;J_i\,.\label{eq:stabilize}
\end{eqnarray}
\fi

To check  this inclusion, one can apply     standard computer
algebra techniques. For example, one can check if $\pi^{(i+1)}[b]\in
J_i$  for each $b\in B_i$, thus solving $|B_i|$ ideal membership
problems, for one and the same ideal $J_i$.  As already discussed, this
presupposes  the computation of a   Gr\"{o}bner basis for $J_i$, a
potentially  expensive operation. One advantage of the above
algorithm, over     methods proposed in program analysis with
termination conditions based on testing ideal membership (e.g.
\cite{Carbonell}),
 is that \iflncs $(\mathbf{*})$ \else\eqref{eq:stabilize}\fi \ is  not checked at every iteration, but only when $V_{i+1}=V_i$
 (the latter a relatively inexpensive check).   %, we need to check if \eqref{eq:stabilize}.
 \iflncs\else
In the Appendix (Subsection \ref{sub:pseudo}), we also discuss a
sufficient  condition which does not involve Gr\"{o}bner bases, but
leads to an incomplete algorithm.
\fi

\begin{exa}\label{ex:running3bis}{%\em
Consider the initial value problem of Example \ref{ex:running0} and
the template $\pi=a_1x+a_2y+a_3z+a_4w$. We run the double chain
algorithm with this system and template as inputs. In what follows,
$v=(v_1,v_2,v_3,v_4)^T$ will denote a generic vector in $\field^4$.
Recall that $\xx=(x,y,z,w)^T$ and $v_0=(0,0,1,1)^T$. {\small
\begin{itemize}
\item For each $v\in \field^4$: $\pi^{(0)}[v](v_0)=(v_1x+v_2y+v_3z+v_4w)(v_0) = 0$ if and only if $v\in V_0\defin \{v\,:\, v_3=-v_4\}$.
\item For each $v\in V_0$: $\pi^{(1)}[v](v_0)=(v_1 x z + v_1 z + v_2wy + v_2 z + v_4w-v_4z)(v_0) = 0$ if and only if $v\in V_1\defin \{v\in V_0: v_1=-v_2\}$.
\item For each $v\in V_1$: $\pi^{(2)}[v](v_0)=(v_2w^2 y + v_2 w y + v_2 w z - v_2 x z^2 - v_2 x z - v_2 z^2 + v_4 w - v_4 z)(v_0) = 0$ if and only if $v\in V_2\defin V_1$.\\
       Being $V_2=V_1$, we also check if   $J_2=J_1$. A basis of $V_1$ is   $B_1=\{b_1,b_2\}$
       with $b_1=(-1,1,0,0)^T$   and $b_2=(0,0,-1,1)^T$. According to \iflncs $(\mathbf{*})$\else\eqref{eq:stabilize}\fi,
       we have therefore to check if, for $\ell=1,2$:
       \iflncs $\pi^{(2)}[b_\ell]   \in   J_1  \defin  \ide{\{\pi^{(0)}[b_1],\pi^{(0)}[b_2],\pi^{(1)}[b_1],\pi^{(1)}[b_2]\}}\,$.
       \else
       {\small
       \begin{eqnarray*}
        \pi^{(2)}[b_\ell] & \in & J_1  \defin  \ide{\{\pi^{(0)}[b_1],\pi^{(0)}[b_2],\pi^{(1)}[b_1],\pi^{(1)}[b_2]\}}\,.
        \end{eqnarray*}}
        \fi
With the help of a computer algebra system, one computes a   Gr\"{o}bner basis for $J_1$ as $G_1=\{x - y,  z-w \}$.
Then one can reduce $\pi^{(2)}[b_1]=w^2 y + w y + wz - x z^2 - x  z - z^2$ modulo $G_1$ and obtain
      $\pi^{(2)}[b_1] = h_1(x - y)+h_2(z-w)$, with $h_1=-z^2 - z$ and $h_2=-wy - yz - y - z$, thus proving that $\pi^{(2)}[b_1]\in J_1$. One proves similarly that
    $\pi^{(2)}[b_2]\in J_1$. This shows that $J_2=J_1$.
\end{itemize}
}
Hence the algorithm terminates with $m=1$ and returns $(V_1,J_1)$, or,
concretely, $(B_1,G_1)$. In particular, $x-y\in \Zzero$, or
equivalently $x(t)=y(t)$. Similarly for $z-w$. }
\end{exa}

\iflncs
 \else
 \begin{rem}[notational convention: result template]\label{rem:convention}{%\em
According to Theorem \ref{th:corr}(a), given a template $\pi$ and $v\in \field^n$, checking if $\pi[v]\in \pi[V_m]$ is
equivalent to  checking if $v\in V_m$, which can be effectively done
 knowing  the basis $B_m$ of   $V_m$ concretely returned by the
 algorithm
(another, equivalent possibility, is checking if $v$ is orthogonal
to the space $V_m^\bot$, which is built in the minimization phase,
see next section).

In practice, it is  more convenient
to represent the whole set $\pi[V]$ returned by the algorithm more  compactly, in terms of a \emph{new}
 $m$-parameters result template $\pi'$ such that $\pi'[\reals^m]=\pi[V]$. For instance, in the previous example, the result template
\begin{eqnarray*}
\pi'  & =  & a_1x-a_1y+a_2z-a_2w
\end{eqnarray*}
represents the algorithm's outcome $\pi[V_2]$, in the precise sense that
 $\pi[V_2]=\pi'[\field^2]$.
}
\end{rem}

\iflncs
\begin{rem}[linear systems]\label{rem:linsys}
When the system of \ode's is linear, that is when the drifts $f_i$ are linear functions of the $x_i$'s, stabilization of the chain of vector spaces can be detected
without resorting to ideals. The resulting single chain  algorithm essentially boils down to the  {`refinement'} algorithm of \cite[Th.2]{Bor09}. See \cite{Full} for details.
\end{rem}
\else
\begin{rem}[linear systems]\label{rem:linsys}{%\em
Consider the case of a \emph{linear} system, that is, when the drifts  $f_i$   in $F$ are linear functions of the $x_i$'s. Consider
 the chain  $V_0\supseteq V_1\supseteq \cdots$ in \eqref{eq:Vi}. It is easy to prove that
as soon as $V_{m+1}=V_m$ then the chain has stabilized, that is $V_{m+k}=V_m$ for each $k\geq 0$. Therefore, for linear systems,
 stabilization can be detected without looking   at the ideals chain \eqref{eq:Ji}, hence dispensing with Gr\"{o}bner bases.
The resulting single chain  algorithm boils down to the  {`refinement'} algorithm of
\cite[Th.2]{Bor09}.}
\end{rem}
\fi

\begin{rem}\label{rem:latnaiv}{%\em
We end this section pointing out that the naive algorithm of
Subsection \ref{sec:algorithm}.1 admits a generalization that works
with templates as well. Specifically,  one can regard   templates as
polynomials in both the variables $\xx$ and parameters $\mathbf{a}$,
and compute the invariant ideal in $\reals[\mathbf{x},\mathbf{a}]$
generated by all Lie derivatives of the given template $\pi$. Having
computed this, one substitutes $v_0$ for $\xx$.  This procedure is
related to what is done by M\"{u}ller-Olm and Seidl in the
discrete-time case \cite{Muller} (further coniderations in the
concluding section).

In any case, the idea of the algorithm in Subsection \ref{sec:algorithm}.2 is to avoid
treating the parameters in the template \emph{symbolically}--- which
is important, bearing in mind that the ideal membership problem
requires exponential space in the number of variables. }
\end{rem}

\section{Minimization}\label{sec:minim}
%Assume one is interested %in simulating a system,   in solving, or
%even in realizing  it. That is,
%in computing, simulating or even realizing the individual functions
%$x_i(t)$ of the trajectory, or polynomial combinations of
%them. %Then it is not immediately clear how the algorithm in the
%preceding section may be useful.
%A method that allows one to transform the original system of
%polynomial \ode's system into a smaller  but still equivalent one,
%is of paramount importance.
We present     a method for reducing the size of   an initial value
problem. \iflncs\else The resulting reduced problem, while
equivalent in a precise sense to the original problem, is
\emph{minimal} in terms of number of equations and variables, among
all   systems that can be obtained by linear aggregations of the
original equations.

\fi
\iflncs
The basic idea is    projecting the original
  system   onto a suitably chosen subspace of $\field^N$. Consider the subspace $W\subseteq \field^N$ of all vectors that are orthogonal to $\xx(t)$ for each $t$ in the domain of definition $D$ of the time trajectory, that is $W=\{v\in \field^N : \langle v\,, \,\xx(t)\rangle =0 \text{ for each }t\in D\}$.
  It is not difficult to prove (see \cite{Full}) that $W=V^\bot_m$,
  where $V_m$ is the subspace returned by the double chain algorithm when fed with
  the linear template $\pi = \sum_{i=1}^N a_ix_i$. Let $B$ the $N\times l$ ($l\leq m+1,N$)
   matrix whose columns are the vectors of   an orthonormal basis of $W$.
   Then,  for each  $t$, $B^T\xx(t)$ are the   coordinates of $\xx(t)$ in the subspace $W$ w.r.t. the chosen basis.
   Consider now  the  following (reduced) initial value problem
   $\redpr$,
    in the new variables $\yy=(y_1,...,y_l)^T$, obtained by projecting the original problem $\ivp$
     onto $W$.
\begin{wrapfigure}{r}{0.35\textwidth}
\vspace*{-1cm}
\begin{equation}\label{eq:exreduced}
\redpr:\;\left\{\begin{array}{rcl}
        \dot\yy(t) & = &  B^TF\left(B\yy(t)\right) \\
        \yy(0) & = & B^T\xx(0)\,.
        \end{array}
\right. \vspace*{-.2cm}
\end{equation}
%\vspace*{-.35cm}
\end{wrapfigure}
In \cite{Full}, we prove the following result. In essence, all
information about $\ivp$ can be recovered exactly from the reduced
$\redpr$, which is the best possible linear reduction of $\ivp$.
%In \cite{Full}, we prove the following result. In essence, all information about $\ivp$ can be recovered exactly from the reduced $\redpr$, which is the best possible linear reduction of $\ivp$.

\begin{thm}[minimal exact reduction]\label{th:exactMinimal}
Let $\yy(t)$ be the unique analytic solution of the (reduced)
 problem \eqref{eq:exreduced}. Then, $\xx(t)=B\yy(t)$. Moreover, suppose for some  $N\times k$ matrix $C$ and vector function
$\zz(t)$, we have $\xx(t)=C\zz(t)$, for each $t\in D$. Then $k\geq l$.
\end{thm}
\else

The method takes as an input the space $V_m$ returned by the double chain algorithm in the preceding
section when fed with a certain linear template. The method
itself is quite simple and only
  relies  on simple linear algebraic operations that can be efficiently automated. %, does not require \emph{per se} any extra computation with Grobner basis

%\subsection{Exact reduction}
%We will show that an exact reduction of the system \eqref{eq:ivp}
%can be obtained
The basic idea is    projecting the original
  system of equations onto a suitably chosen subspace of $\field^N$.
 Consider the
linear  template
\begin{eqnarray}\label{eq:lintempl}
\pi& =& a_1\cdot x_1+\cdots + a_N\cdot x_N%\sum_{i=1}^N a_i x_i
\end{eqnarray}
where the $a_i$'s are  distinct parameters. By applying the
algorithm of the preceding section to this template, we obtain a
subspace $V\defin V_m\subseteq \field^N$. Consider now the orthogonal
complement of $V$ in $\field^N$ (where $\langle\cdot,\cdot\rangle$
is the usual   scalar product between vectors in $\field^N$ and
$v,w$ denote    generic vectors in $\field^N$)
\begin{eqnarray*}
W& \defin & V^\bot \;=\;\{w\in \field^N\,: \,\langle w, v\rangle
=0\text{ for each } v\in V\}\,.
\end{eqnarray*}
We show that the trajectory $\xx(t)$ lies entirely in $W$, that is
$\xx(t)\in W$ for each $t$ in the open interval of definition, say
$D$, of the trajectory. Indeed,
 by virtue of Theorem \ref{th:corr}(a),    we have that $v\in V$ if and only if (here $v=(\lambda_1,...,\lambda_N)^T$)
 \[
\begin{array}{rclclcl}
\langle \xx, v\rangle & =  &  \sum_{i=1}^N \lambda_ix_i     &  = & \pi[v]  & \in &  \Zzero\,. %& \text{ if and only
%if } & v\in V\,.
\end{array}
\]
By definition of $\Zzero$, this means that $v\in V$ if and only if
 \[
\begin{array}{rclclcl}
\langle \xx(t), v\rangle & = & \sum_{i=1}^N \lambda_ix_i(t) &  = & \pi[v](t)      & = &   0  \;\; \text{ for each }t %& \text{ if and only
%if } & v\in V\,.
\end{array}
\]
that is $\xx(t)\in W$ for each $t$ in the open interval $D$ of definition of $\xx(t)$. In other words, $v\in V$ if and
only if $v$ is orthogonal to each $\xx(t)$, for $t\in D$, or $V=\{\xx(t):t\in
D\}^\bot$. This is equivalent to the following crucial lemma.
%, in view of the above equivalence,
%we have the stronger result that
%: as  $v\in V$ if and only if $\langle\xx(t),v\rangle=0$ for
%each $t\in D$  is equivalent to
%$V=\{\xx(t):t\in D\}^\bot$, or
%equivalently, we have the following.

\begin{lem}\label{lemma:spanxt}
$W=V^\bot=\myspan\{\,\xx(t):t\in D\,\}$.
\end{lem}

The fact that the trajectory $\xx(t)$ entirely lies in, and in fact generates,  the subspace
$W$, suggests that we can obtain a more economical representation of
$\xx(t)$ by adopting a system of  coordinates in this subspace. More formally, let $B$ be any orthonormal basis of $W$. It
is convenient to represent $B$  as a matrix of $l$ independent
column vectors, $B=[b_1|\cdots |b_l]\in \field^{N\times l}$, where
$l\defin \dim(W)\leq N$ (in fact, $l\leq m+1$ as well; \iflncs see \cite{Full}\else  this is
discussed in the Appendix, Subsection \ref{sub:basis}, where  an
efficient method  for  building $B$ out of the successive Lie
derivatives of $\pi$ is explained\fi). Recall that orthonormality means
$B^TB=I_l$, with $I_l$   the $l\times l$ identity matrix. The
orthogonal projection of any
  $v$ onto $W$ has, w.r.t. $B$, coordinates   $B^T v$, which is of course a vector in  $\field^l$.
 Define now
\begin{eqnarray}\label{eq:yy}
\yy(t) & \defin & B^T\xx(t) \;\; \text{ for each }t\in D\,.
\end{eqnarray}
Since each $\xx(t)\in W$, each $\xx(t)$ is a fixpoint of the
projection, so with the above definition we have
\begin{eqnarray}
\xx(t) & = & BB^T\xx(t)\nonumber\\
       & = & B\yy(t)\,.\label{eq:xproj}
\end{eqnarray}
From the last equation, it is easy to check that $\yy(t)$ is a
solution, hence the \emph{unique} analytic solution in a suitable
interval, of the following \emph{reduced}  problem $\redpr=(G,\yy(0))$,
where $F$ denotes the vector field of the original initial value
problem:
\begin{equation}\label{eq:exreduced}
\redpr:\;\left\{\begin{array}{rcl}
        \dot\yy(t) & = &  B^TF\left(B\yy(t)\right) \\
        \yy(0) & = & B^T\xx(0)\,.
        \end{array}
\right.
\end{equation}
In order to check that $\yy(t)$ as defined in \eqref{eq:yy}
satisfies the first equation of \eqref{eq:exreduced}, observe that,
by definition   we have:
\begin{eqnarray*}
\dot\yy(t) & = & B^T\dot\xx(t)\\
           & = & B^TF\left(\xx(t)\right)\\
           & = & B^TF\left(B\yy(t)\right)
\end{eqnarray*}
where the last equality follows from \eqref{eq:xproj}. The second
equality of \eqref{eq:exreduced} is trivially seen to be true. Note
that the reduced system \eqref{eq:exreduced}  features  $l\leq N$
differential equations. In particular,    observe that the vector
field $G$ of the reduced system  is obtained   by replacing
each variable $x_i$ in the original $F$
 with a linear combination of the variables $y_j$' s, as dictated by $B$, and
then linearly aggregating  the resulting $N$ terms, as dictated by
$B^T$. As a consequence,  the maximum degree in the reduced $G$
  does not exceed the maximum degree in the original $F$.

Equation \eqref{eq:xproj} naturally extends to any polynomial
behaviour. That is, for any polynomial $p\in \field[\xx]$, we have
$p(\xx(t))=p\left( \xx(t)\right)= p\left(B\yy(t)\right)$. In the
end, we have proven the following result, which shows that we can
exactly recover  any behaviour induced by the original system from
the reduced system.

\begin{thm}[exact reduction]\label{th:exred}
Let $\yy(t)$ be the unique analytic solution of the (reduced)
 problem \eqref{eq:exreduced}. Then, $\xx(t)=B\yy(t)$.
Moreover, for any polynomial behaviour $p(\xx(t))$ induced by the
original problem \eqref{eq:ivp}, we have
$p(\xx(t))=p\left(B\yy(t)\right)$.
\end{thm}

The reduced system is minimal among all  systems obtained as linear
aggregations of the original  system. In the next result, the
interpretation of the $k$-dimensional vector function $\zz(t)$ is
that it may (but need not) arise as the solution of any system with
$k$ equations.

\begin{thm}[minimality]\label{th:minimal}
Assume for some  $N\times k$ matrix $C$ and vector function
$\zz(t)$, we have $\xx(t)=C\zz(t)$, for each $t\in D$. Then $k\geq l$.
\end{thm}
\begin{proof}
Assume $k\leq N$ (otherwise there is nothing to prove). As the
trajectory $\xx(t)$ spans $W$ (Lemma \ref{lemma:spanxt}), which has
dimension $l$, we can form a rank $l$ matrix $E$ as
$E=[\xx(t_1)|\cdots |\xx(t_l)]= [C\zz(t_1)|\cdots |C\zz(t_l)]=
C[\zz(t_1)|\cdots |\zz(t_l)]$, for suitable points $t_1,...,t_l$. As
in general $\rk(AB)\leq \min\{\rk(A),\rk(B)\}$, we have $\rk(C)\geq
l$. But $k\geq \rk(C)$, which implies the thesis.
\end{proof}

As a corollary of Theorem \ref{th:exred}, we obtain a further
characterization of \lbisim-bisimilarity, which shows that we can
also reason syntactically on polynomial behaviours in terms of the
reduced system. In particular, \lbisim-bisimilarity between pairs of
individual variables reduces to plain equality between the
corresponding rows of $B$, which allows one to easily form
equivalence classes of variables if desired.  Let us denote by
$G=(g_1,...,g_l)^T$ the polynomial vector field of the reduced
system. Note that $G$ is expressed in terms of the variables
$\yy=(y_1,...,y_l)^T$, more precisely $G=B^T F(B\yy)$.

\begin{cor}\label{cor:redbisim}
Let $p,q\in\field[\xx]$. Then $p\sim_{\ivp} q$ in $C_\ivp$ if and only if
$p(B\yy) \sim_{\redpr} q(B\yy)$ in $C_\redpr$. In particular, $x_i\sim_\ivp x_j$
in if and only if row  $i$ equals row $j$ in $B$.
\end{cor}

\begin{exa}\label{ex:running4}{%\em
Let us consider again the problem of Example \ref{ex:running1}.
Recall from Example \ref{ex:running3bis} that the algorithm for
computing invariants
stops with $m=1$ returning $V_1,J_1$. %, with
%Here $\xx=(x,y,z,w)^T$. Consider the linear template $\pi=a_1x+a_2y+a_3z+a_4w$.
%When run with this template, the algorithm  stops with $m=2$ and yields the constraints $a_1=-a_2$ and $a_3=-a_4$, which correspond to the 2-dimensional
% space
In particular, $V\defin V_1=\myspan\{(1,-1,0,0)^T,\,(0,0,1,-1)^T\}$.
It is easily checked
that %an orthonormal basis for
$W=V^\bot=\myspan\{c_1,c_2\} $, where $c_1=(1/\sqrt 2,1/\sqrt
2,0,0)^T$ and $c_2=(0,0, 1/\sqrt 2,1/\sqrt 2)^T$ are orthonormal
vectors.}
%\end{exa}
%
%\begin{wrapfigure}{r}{0.34\textwidth}
%\vspace*{-.4cm}
%\begin{equation*}
%\left\{\begin{array}{rcl}
%        \dot y_1(t) & = &  \frac 1{\sqrt 2}y_1(t)y_2(t)+y_2(t)\\
%        \dot y_2(t) & = & y_2(t)\\
%        \yy(0) & = & (0,\sqrt{2})^T\,.
%       \end{array}
%\right.
%\end{equation*}
%\vspace*{-.35cm}
%\end{wrapfigure}
%
%\begin{equation*}%\label{eq:exreduced}
%\begin{figure}
%
%
%\vspace*{-.4cm}
Hence we let the basis matrix be $B=[c_1|c_2]$.
By applying \eqref{eq:exreduced}, we build the minimal system in the
variables $\yy=(y_1,y_2)^T$ shown below.
\begin{equation*}
\left\{\begin{array}{rcl}
        \dot y_1(t) & = &  \frac 1{\sqrt 2}y_1(t)y_2(t)+y_2(t)\\
        \dot y_2(t) & = & y_2(t)\\
        \yy(0) & = & (0,\sqrt{2})^T\,.
        \end{array}
\right.
\end{equation*}
Note that the first and second rows of $B$ are equal, as well as the
third and the fourth, which proves (again)   that $x(t)=y(t)$ and
$z(t)=w(t)$.
\end{exa}
\fi

%\vspace*{-.31cm}
%\hspace*{-.3cm}
%\end{figure}
%\caption{\label{fig:plotx}This is a figure caption.}

%\begin{eqnarray*}
%(\pi[v])(t)=\sum_{i=1}^N v_ix_i(t)\; =\; 0 & \text{ if and only if } & v\in V\,.
%\end{eqnarray*}
%As seen in Subsection \ref{sub:eff}, the space $V$ can be concretely
%characterized as the null space of a suitable matrix $T_m=T\in
%\field^{m\times N}$, that is $V=\ker(T)$, where the matrix $T$ can
%be built incrementally row by row, as the Lie derivatives of $\pi$
%are generated.

%\ifmai
\section{Approximate reduction and linearization}\label{sec:approx}
If one is ready to accept some       degree of
approximation, it is possible to build a
reduced system   in a simpler,   linear  form. %We tackle these
%issues in the following.
%Exact  and approximate reduction will be
%dealt with separately in the following.
%
To this purpose, we will illustrate a method for approximate  linearization. The
polynomial behaviours induced by the original and by the reduced
system   will in general differ by a term which is $O(t^{m})$,
for some prescribed $m$. This may be useful for analysis of the
beaviour of the system  around a chosen operation point $t_0$, which
here will be conventionally fixed as $t_0=0$. The approximation can
of course be very bad for $t$ too far from $0$ (see  in the concluding section
the discussion  on {\sc tpwl} for a strategy    to possibly recover global accuracy).
The method is based entirely on simple linear algebraic
manipulations. Another attractive feature  is that the reduced
system, besides being linear,  is  quite small, featuring no more
than $m$ equations. We will also give conditions under which the
reduction is exact.

Let $S\subseteq \field[\xx]$ be a set of polynomials of bounded
degree -- that is,  there is $k$ such that $\deg(p)\leq k$ for each
$p\in S$. For example, one might take $S=\pi[\field^n]$, for a
certain template $\pi$ with $n$ parameters. Assume we are interested
in computing/simulating  the set of functions $\{p(\xx(t))\,:\,p\in
S\}\subseteq \A$ and are ready to accept an approximation error
around 0  of order $m$, for a fixed  $m\geq 1$.

The basic idea
 is viewing  $\lie$ as a linear operator on the vector space of polynomials, and then taking its projection onto a small,
 suitably chosen subspace.  Let $\M=\{\alpha_1,...,\alpha_M\}$ be the set of     distinct monomials appearing
 in the polynomials of $\cup_{j=0}^{m-1}\lie^{(j)}(S)$. %, totally ordered in some way.  %integer  large enough that $p^{(j)}\in \field_d[\xx]$ for all $p\in V$ and $0\leq j\leq m-1$.
%This $d$ exists and can be statically computed from the vector field $F$, provided the degree of the polynomials in $V$ is bounded.
%Now fix any total  order of   the distinct monomials   in $\field_d[\xx]$,    as  $\alpha_1,...,\alpha_M$:
 Let $\UM$ be the set of  polynomials that can be generated starting from $\M$. Clearly, $\UM\subseteq
\field_d[\xx]$  for some $d$ large enough. Moreover, when we regard
polynomials  as (column) vectors whose components are indexed by
monomials, totally ordered in some way,  $\UM$ is isomorphic to
$\field^M$  as a $M$-dimensional  vector space over the field
$\field$. Let $\pro_{\UM}$ be the orthogonal projection from
$\field[\xx]$  onto $\UM$, and consider   the function
$\pro_{V}\circ \lie:\UM\rightarrow  \UM$.  We denote by $L$   the
$M\times M$ matrix representing $\pro_{\UM}\circ \lie$ w.r.t. the
canonical basis of $\UM$, that is $\M$. By definition,
$(\pro_{\UM}\circ\lie)_{|S}=\lie_{|S}$.  This implies that, for each
$p\in S$,    $\lie(p)=Lp$, and more generally that
\begin{eqnarray}\label{eq:linearLie}
\lie^{(j)}(p) & = & L^j p \ \ \ \text{ ($p\in S$ and $0\leq j\leq
m-1$).}
\end{eqnarray}
Moreover, if we   evaluate the given monomials at $v_0$ and let
\begin{eqnarray*}
\aalpha^T & \defin & (\alpha_1(v_0),...,\alpha_M(v_0))
\end{eqnarray*}
then
\begin{eqnarray}\label{eq:valueLie}
(\lie^{(j)}(p))(v_0) & = & \phi^T L^j p \ \ \ \text{{\small ($p\in
S$ and $0\leq j\leq m-1$).}}
\end{eqnarray}
Now consider   the   subspace of $K_m$ defined as follows
\begin{eqnarray}\label{eq:krylov}
K_m & \defin & \myspan \left\{\aalpha, L^T\aalpha,...,
(L^T)^{m-1}\aalpha\right\}\,.
\end{eqnarray}
This is the order-$m$ \emph{Krylov space} generated by the matrix $L^T$ and by $\aalpha$. %In  what follows, we assume that the $m$ vectors above are linearly independent.
%
%For any subspace $U$, denote by $\pr_{U}$ the orthogonal projection from $\field_d[\xx]$ onto $U$, and by $U^\bot$ the orthogonal complement of $U$ in $\field_d[\xx]$.

%\begin{lem}\label{lemma:annihil} Let $p,q\in V$. The following three statements are equivalent: (a) $p(\xx(t))=_m q(\xx(t))$ in $\A$; (b) $\pr_{K_m}(p)=  \pr_{K_m}(q)$ in $K_m$; (c) $p-q\in K^\bot_m$.
%\end{lem}
%\begin{proof}
%Note that, for each $0\leq j\leq m-1$
%\begin{eqnarray}
%p^{(j)}(0)& = &  (\lie^{(j)}(p))(0)\label{eq:step1}\\
%          & = & \aalpha^T L^j p\\
%          & = & p^T (L^T)^j \aalpha
%\end{eqnarray}
%where \eqref{eq:step1} is a consequence of \eqref{eq:lieder}. Similarly for $q$. As a consequence, $p^{(j)}(0)=q^{(j)}(0)$ if and only if $(p-q)^T (L^T)^j \aalpha =0$. From this, it is immediate to check that $p(\xx(t))=_m q(\xx(t))$ if and only if $p-q$ is orthogonal to $K_m$, which proves the equivalence between (a) and (c). The equivalence between (c) and (b) is obvious.
%\end{proof}

Fix now any orthonormal basis of $K_m$, which we may represent as a
$M\times l$  matrix $B=[b_1|...|b_{l}]$, for some $l\leq m$
(orthonormality means $B^TB=I_{l}$). The orthogonal projection from
$\UM$ onto $K_m$ is therefore given by $\pro_{K_m}(v)=BB^T v$.
%\mik{Is orthonormality needed?}
Consider the function that, taken a vector $v\in \UM$, applies $L^T$
and then projects onto $K_m$
\begin{eqnarray*}
v\mapsto \pro_{K_m}(L^Tv)= BB^TL^Tv\,.
\end{eqnarray*}
When restricted to $K_m$,  this defines a linear morphism
$K_m\rightarrow K_m$.  Call $A$ the $l\times l$ matrix representing
this morphism w.r.t. the basis $B$. Explicitly,
\begin{eqnarray*}
A & = & B^TL^TB\,.
\end{eqnarray*}

\begin{lem}\label{eq:Aj} For $0\leq j\leq m-1$, we have $(L^T)^j \aalpha = BA^jB^T \aalpha$.
\end{lem}
\begin{proof}
This is an easy consequence of the fact that, for any $v\in K_m$,
$v=BB^Tv$, and  of the definition of $A$.
\end{proof}

Let us now introduce a new vector of $l$ variables, $\yy= (y_1,...,y_{l})^T$ %.
%We will consider the polynomials in $\yy$ $\field[\yy]$ as isomorphic
and consider the following   \emph{linear} initial value problem,
given by the  matrix $A$
\begin{equation}\label{eq:reduced}
\left\{\begin{array}{rcl}
        \dot\yy(t) & = &  A \yy(t) \\
        \yy(0) & = & B^T\aalpha\,.
        \end{array}
\right.
\end{equation}
Let us denote by $\yy(t)=(y_1(t),...,y_{l}(t))^T$ the (unique)  analytic   solution of this system. Recall that, given $p\in \field[\xx]$, we let $p(\xx(t))\in \A$ denote the function which extends $p\circ \xx(t)$. The following theorem says that, given $p\in S$,  we can reconstruct $p(\xx(t))$, within an approximation of order $m$, by taking a linear combination of the $y_i(t)$ s, with coefficients  given by projecting $p$ (as a vector in $\UM$) onto $K_m$. % and   $r(\yy(t))\circ \yy(t)$ to denote.
%Given two analytical functions $f(t),g(t)\in \A$, we will write $f(t)=_m g(t)$ if $f(t)-g(t)=O(t^{m})$.

\begin{thm}\label{th:pt} Let $p\in S$ and let $\yy(t)$ be the unique   solution of the (reduced)
problem \eqref{eq:reduced}. Then $p(\xx(t)) - p^TB\yy(t) = O(t^m)$
as $t\rightarrow 0$.
\end{thm}
\begin{proof} By definition, the  derivatives of $\yy(t)$ from  the $0$-th through the $(m-1)$-th,  evaluated at $t=0$,  can be written as follows:
{\small
\begin{eqnarray*}
\yy(0) & = & B^T\aalpha\\
\yy^{(1)}(0) & = & AB^T\aalpha\\
 & \vdots & \\
\yy^{(m-1)}(0) & = & A^{m-1}B^T\aalpha\,.
\end{eqnarray*}}
Therefore, for $j=0,...,m-1$, we have
%{\small
\begin{eqnarray}
 p^{(j)}(0) & = & (\lie^{(j)}(p))(\xx(t))_{|t=0}\nonumber\\
            & = & (\lie^{(j)}(p))(v_0)\nonumber\\
            & = & \aalpha^T L^j p\label{eq:pt0}\\
            & = & p^T(L^T)^j\aalpha\nonumber\\
            & = & p^T BA^jB^T\aalpha\label{eq:pt2}\\
            & = & p^T B\yy^{(j)}(0)\label{eq:pt3}\\
            & = & \frac{d^j}{dt^j}\left((p^T B)\yy(t) \right)_{|t=0}
\end{eqnarray}%}
where:   \eqref{eq:pt0} is  \eqref{eq:valueLie},   \eqref{eq:pt2}
follows from  Lemma \ref{eq:Aj}, \eqref{eq:pt3} from the above
expressions for the derivatives of $\yy(t)$ at 0. This way, we have
proved that the first $m$   coefficients in the Taylor expansion
around 0 of $p(\xx(t))$ and  $(p^TB)\yy(t)$ are the same, which is
the wanted statement.
\end{proof}

\begin{figure}[t]
\label{fig:plots}
\begin{center}
\includegraphics[width=0.5\textwidth]{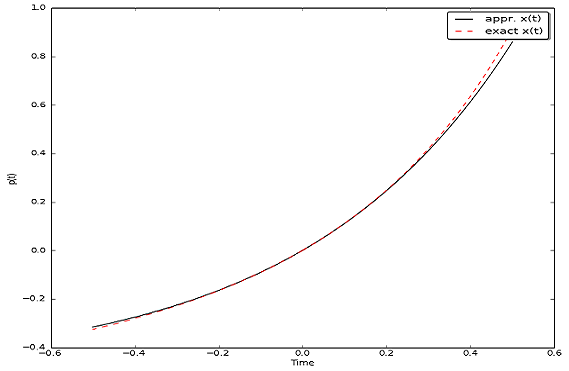}
\end{center}
\caption{Plots of exact and approximate solutions.}
\end{figure}

\begin{exa}\label{ex:running6}{%\em
Consider again the system of
Example \ref{ex:running1}. Assume we are interested in the
  behaviours described by $S=\{x,y,z,w\}$. The following
reduced linear system in the variables $\yy=(y_1,y_2,y_3)^T$, which
guarantees an approximation of order $m=3$ for those behaviours, can
be obtained by applying the  method.
%{\small
\begin{equation}%\label{eq:reduced}
\left\{\begin{array}{rclclcl}
        \dot y_1 & = &  3y_1/2 & + & \sqrt 5 y_2/10 &+ & \sqrt{30}y_3/30 \\
        \dot y_2 & = &  \sqrt 5 y_1/2 &+ &   11 y_2/10 &+ &
         \sqrt 6 y_3 11/30\\
        \dot y_3 & = &         &        & \sqrt 6 y_2/5 &+&  4
        y_3/10\\
        \yy(0) & = &\multicolumn{5}{l}{(2,0,0)^T\,.}
        \end{array}
\right.
\end{equation}
%
%
%\begin{wrapfigure}{r}{0.25\textwidth}
%\vspace*{-.31cm}
%\hspace*{-.3cm}
%\vspace*{-.4cm}
%\caption{\label{fig:plotx}This is a figure caption.}
%\end{wrapfigure}
%
We do not report the full $4\times 3$ matrix $B$ here, but only
mention that, following Theorem \ref{th:pt},  the approximated
version of $x(t)$ can be obtained from $\yy(t)$ as: $\hat
x(t)=(1,0,0,0)\cdot B\yy(t)=\sqrt 5 y_2(t)/5 - \sqrt{30}y_3(t)/10$.
The plots in Figure \ref{fig:plots} show that the approximation is quite good
around $t=0$.
}
\end{exa}

We finally give a sufficient condition for exactness. This
 condition applies, for example, to linear systems, that is when the drifts $f_i$  in $F$ are linear functions.
    Consider the sequence of Krylov spaces $K_1,K_2,...$ that can be built according to equation \eqref{eq:krylov}. A Krylov space $K_m$  is said to
    be \emph{invariant} if $K_{m+1}\subseteq K_m$. Of course there always exists
     $m\leq M$ such that $K_m$ is invariant;
     moreover the condition of invariance for $K_m$ can be efficiently detected (see \cite{Saad}).

\begin{cor}\label{cor:exact}
Assume $\M$ includes all monomials in  $\cup_{j\geq 0}
\lie^{(j)}(S)$. Choose $m$ such that   $K_m$ is invariant. Then,
with the  notation introduced above, $p(\xx(t)) = p^TB\yy(t)$.
\end{cor}
\begin{proof} Under the given condition, $\lie^{(j)}(p)=Lp$ for any $j\geq 0$ and $p\in S$.
Moreover, since $K_m=K_{m+1}=\cdots$, the statement in Lemma
\ref{eq:Aj} also holds for any $j\geq m$, hence for any $j$. This
allows one to extend the proof given in \ref{th:pt} to prove  that
the Taylor coefficients of
 $p(\xx(t))$ and $(p^TB)\yy(t)$ are all
the same.
\end{proof}

\begin{rem}[on computing the matrices $A$ and $B$]\label{rem:Arnoldi}{%\em
There exist relatively efficient and  numerically stable methods to
build the pair of matrices $A$ and $B$ needed in the construction of
the reduced system. One of these methods is the \emph{Arnoldi
algorithm} \cite{Arnoldi,Saad}, which takes $O(Nz\cdot m)$
floating-point operations, and as much memory if a sparse storage
scheme is adopted. Here $Nz$ denotes the number of nonzero elements
in the matrix $L$: this is $O(M^2)$ in the worst case, but it will
be typically much smaller, as polynomials arising in applications
tend to be sparse. Moreover, the method need not a fully stored
matrix $L$, but only  a handle to the matrix-vector multiplication
function $u\mapsto L^Tu$. Using appropriate data structures, this
method can be implemented according to an `on-the-fly' strategy,
 where the terms are unfolded as needed. %See \cite{Bor09,Bor15} for further details on this aspect.
}
\end{rem}
%\fi

\section{Examples}\label{sec:examples}
\iflncs We have put a proof-of-concept
implementation\footnote{Python code   available from \url{...}.
Reported execution times relative to
 the pypy interpreter under Windows 8 on a core i5 machine.}  of our algorithms to work on
a few simple examples taken from the literature.
We illustrate
below  two cases.
\else
Although the focus of the present paper is mostly on theory, it is instructive to
 put a proof-of-concept implementation\footnote{Python code
 available at \url{http://local.disia.unifi.it/boreale/papers/DoubleChain.py}. Reported execution times relative to
 the pypy interpreter under Windows 8 on a core i5 machine.}  of our algorithms at work on
a few simple examples taken from the literature.
We illustrate
below  two cases,     a linear and a nonlinear system. %Wherever appropriate, we also
%compare our results to those obtained with the Erode tool by
%Cardelli et al. \cite{Erode}. This will help to highlight the main
%differences between our approach and theirs.
%The implementation  and examples are available from the author.
\fi

\ifmai
\subsection{Example 1: complexation}\label{sub:decomp}
The following system, which we have taken from them documentation of
\cite{Erode},
describes a  reaction involving    two reactants, $A$ and $B$.
 The first of the two may be in either   of  two internal states (phosphorylated $A_p$ / unphosphorylated $A_u$), which binding with $B$
  may form, respectively
 compounds $C_p$ and $C_u$.  A compound can in turn be decomposed into the original components.
  The equations below describe the time evolution of the concentration of the species in this system,
  starting from an initial condition. Here,   $\xx=(A_u,A_p,B,C_u,C_p)^T$.  % and $v_0=(1,1,3,0,0)$.
{\small
\[
\left\{
\begin{array}{rcl}
\dot{A_u}  & = & -A_u + A_p - 3A_u\cdot B + 4C_u\\
\dot{A_p} & = &  A_u - A_p - 3A_p\cdot B + 4C_p\\
\dot{B} & = & -3A_u\cdot  B + 4 C_u - 3 A_p\cdot B + 4C_p\\
\dot{C_u} & = & 3 A_u \cdot  B - 4 C_u\\
\dot{C_p} & = & 3 A_p \cdot  B - 4 C_p\\
\xx(0) & = & (1,1,3,0,0)^T\,.
\end{array}
\right.
\]
} When run    with this system and the linear template $\pi$ in
\eqref{eq:lintempl} as inputs,
 the double chain algorithm of Subsection \ref{sub:algo} terminates  after $m=2$ iterations
 (in about 4 seconds\footnote{Execution times for a  core i5 machine under Windows 8.1.}, with the
  pseudoideal heuristics of
    Subsection \ref{sub:pseudo}),
% with $\pi[V_4]=    a_1(P  + S - E -C_2)$, for a parameter $a_1$, using the notational convention of Subsection \ref{sub:templ}.
 %The algorithm terminates with $m=2$ iterations (in about 4 seconds, with the heuristics not relying on   Gr\"{o}bner bases,
 % cf. Subsection \ref{sub:pseudo}),
  returning a basis for each of $V_2,J_2$. In particular, for real parameters $a_1,a_2,a_3$, we
  have, with the notational convention of Remark
  \ref{rem:convention}, that $\pi[V_2]$ is described by the 3 parameters template
{\small
  \begin{equation*}
  %\pi[V_2] & =  &
  a_2  3 A_u   + a_3   3A_p -(a_2 + a_3)  B  +   a_1  3 C_u  + (-3a_1 + a_2 + a_3)  C_p  \,.
  \end{equation*}}
Therefore    $V_2$ is a 3-dimensional space. We apply the
minimization procedure of Section \ref{sec:minim}:
     $W=V^\bot_m$   is necessarily a 2-dimensional space, meaning that the original system can be reduced to
     \emph{two} equations.
     Explicitly, an orthonormal  basis for $W$ is $B=[b_1|b_2]$ with $b_1=(1, 1, 3, 0, 0)^T \sqrt{11}/11$ and
      $b_2=(-1/44, -1/44, 1/66, 1/12, 1/12)^T \sqrt{66}$.
       Note that the first two rows of $B$ are the same, as well as the last two,
        implying $A_u(t)=A_p(\xx(t))$ and $C_u(t)=C_p(\xx(t))$ (this might have also been deduced by setting $a_1=0,a_2=1,a_3=-1$ and
        $a_1=1,a_2=0,a_3=0$, respectively, in the above polynomial template, of course). The minimal system
        in the variables $\yy=(y_1,y_2)^T$  is, according to
        \eqref{eq:reduced}
{\small
\[
\left\{
\begin{array}{rcl}
\dot{y}_1  & = & -72\sqrt{ 11}y_1^2/121 + 14\sqrt{66} y_1 y_2/121 +
6\sqrt{11}y_2^2/121\\
 && + 8\sqrt 6 y_2/3\\
\dot{y}_2 & = &  18 \sqrt{66} y_1^2/121 - 21\sqrt{11}y_1 y_2/121 - 3\sqrt{66}y_2^2/242\\
&& - 4y_2\\
\yy(0) & = & (\sqrt{11},0)^T
\end{array}\right.
\]
}
%with the initial conditions $\yy_0=(\sqrt{11},0)^T$.
The invariant ideal $J_2$, not reported here,  may also provide  (a
biologist) with useful
 information about nonlinear conservation laws, such as  the following: $(A_p+A_u) B = 2C_p\cdot A_u +6 A_u^2$.

By comparison, the Erode tool   \cite{Erode} finds
a reduced system of \emph{three} equations. We refer here to
reductions obtained via their
  Backward Differential Equivalence (\bde), which is the one that can be directly compared to ours.
  In particular,
while aggregating $A_u,A_p$ and $C_u,C_p$ in the same classes,
\bde\  cannot find, for example, the conservation law $  3 A_u   +
C_p  =  B$, which can be deduced from our template above setting
$a_1=0,a_2=1,a_3=0$.

\subsection{Example 2: competitive inhibition}\label{sub:linear}
We consider a reaction where a substrate $S$ can be transformed into
a compound $C_1$ and then into a product $P$. Both reactions are
catalyzed by the presence of an enzyme $E$. An inhibitor $I$
competes with $S$ in the  binding with $E$ for producing another
compound, $C_2$. Here, $\xx=(S,E,C_1,C_2,I,P)^T$ and  the equations
are {\small
\[
\left\{\begin{array}{rcl}
    \dot S   &  = & -S\cdot E + C_1\\
  \dot E &  = & -S\cdot E -I\cdot  E +2C_1+2C_2\\
  \dot C_1 & =  & S\cdot  E -2C_1\\
  \dot C_2  & = & E\cdot   I-2C_2\\
  \dot I  &  = &-2I\cdot  E +2C_2\\
  \dot P & =  & C_1\\
  \xx(0) & = & (1,1,1,1,1,1)^T\,.
  \end{array}\right.
\]
} In this case the double chain algorithm terminates (in about 0.5
s)  after $m=4$ iterations, returning
 $\pi[V_4]=    a_1(P  + S - E -C_2)$, for a parameter $a_1$.
In other words, the algorithm has found the single conservation law
$P+S=E+C_2$. Since $V_4$ is  a 1-dimensional space,   $W=V_4^\bot$
is 5-dimensional. As a consequence, the minimal system, which we do
not report here, has 5 equations. By comparison, Erode \cite{Erode}
finds no reductions for this system. \fi

\subsection{Example 1: linearity and weighted automata}\label{sub:linear}
%The purpose of this example is to illustrate the difference between
%\bde\ and   \lbisim-bisimilarity,   concerning  the issue of
%branching vs. linear time (or a continuous analog of it).
%
%
The purpose of this example is to   argue that, when the transition
structure induced by the Lie-derivatives is considered,
%\bde\
%is fundamentally branching-time, whereas
\lbisim-bisimilarity is fundamentally a
linear-time semantics. We first introduce \emph{weighted automata},  a more
compact\footnote{At least for linear vector fields, the resulting
weighted automaton is finite.}
 way of representing the transition structure of the coalgebra $C_\ivp$.
%

%\vspace*{-.5cm}

A (finite- or infinite-state)
 {weighted automaton} is like an ordinary automaton, save that both
states and transitions are equipped with   \emph{weights} from
$\field$. Given $F$ and $v_0$, we can build a weighted automaton
with monomials as states, weighted transitions given by the rule
$\alpha\goto{\lambda}\beta$ iff $\lie_F(\alpha)=\lambda\beta+q$ for
some
%\footnote{Here we see polynomials as linear combinations of
%monomials, with $\beta$ not occurring in $q$.}
 polynomial $q$ not comprising $\beta$ as a monomial, and real
$\lambda\neq 0$, and where each state $\alpha$ is assigned weight
$\alpha(v_0)$. As an example, consider the weighted automaton in
Figure \ref{fig:wa}, where the state weights (not displayed) are 1
for $x_{10}$, and 0 for any other state.
\begin{figure}[t]
\ifmai
{\small
\begin{center}
\begin{picture}(111,81)(0,-81)
\gasset{AHnb=1,AHdist=1.41,AHangle=20,AHLength=3.5,AHlength=3.41}
%\put(0,-81){\framebox(111,81){}}
\node[linewidth=0.5](n0)(28.0,-8.0){$x_1$}

\node[linewidth=0.5](n1)(28.0,-28.0){$x_2$}

\node[linewidth=0.5](n5)(88.0,-8.0){$x_5$}

\node[linewidth=0.5](n6)(72.0,-28.0){$x_6$}

\node[linewidth=0.5](n7)(104.0,-28.0){$x_7$}

\node[linewidth=0.5](n8)(72.0,-44.0){$x_8$}

\node[linewidth=0.5](n9)(104.0,-44.0){$x_9$}

\node[linewidth=0.5](n10)(56.0,-64.0){$x_{10}$}

\node[linewidth=0.5](n13)(16.0,-44.0){$x_3$}

\node[linewidth=0.5](n15)(40.0,-44.0){$x_4$}

\drawedge[linewidth=0.5](n0,n1){1}

\drawedge[linewidth=0.5](n1,n13){$2/3$}

\drawedge[linewidth=0.5](n1,n15){$1/3$}

\drawedge[linewidth=0.5](n5,n6){$1/2$}

\drawedge[linewidth=0.5](n5,n7){$1/2$}

\drawedge[linewidth=0.5](n6,n8){1}

\drawedge[linewidth=0.5](n7,n9){1}

\drawedge[linewidth=0.5](n8,n10){1}

\drawedge[linewidth=0.5](n15,n10){$3/2$}

\drawedge[curvedepth=-8.0,linewidth=0.5](n13,n10){$3/4$}

\drawedge[curvedepth=8.0,linewidth=0.5](n9,n10){1}

\drawloop[loopdiam=6.0,loopangle=-90.0,linewidth=0.5](n10){ 1}
\end{picture}
\end{center}
}
\fi
%\centering
\vspace*{-3.5cm}
\begin{center}\hspace*{-.8cm}
\includegraphics[width=0.8\textwidth]{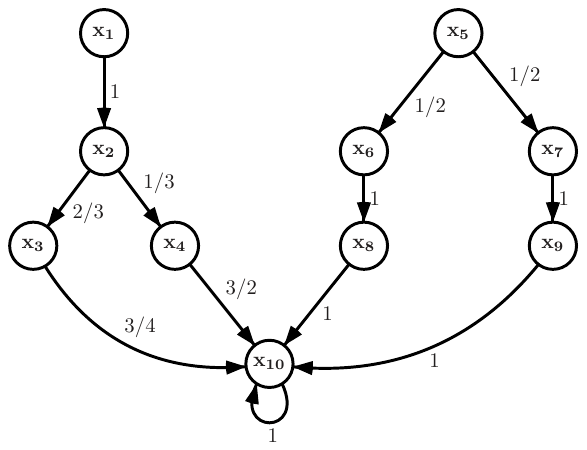}
\end{center}
\vspace*{-9cm} \vspace*{-1cm}
\caption{A weighted automaton.}
\label{fig:wa}
\end{figure}
This automaton is generated -- and in fact codes up -- a   system of \ode's
  with ten variables,  where
    $\dot x_1 = x_2$, $\dot x_2 = (2/3)x_3+(1/3)x_4$
etc.,   with the initial condition as specified by the state weights
($x_1(0)=0$ etc.).
\iflncs The standard linear-time semantics of weighted automata (see \cite{Rutten,IC12} and references therein)
 is in full agreement with $\lie$-bisimilarity \cite{Full}. As a consequence,  in our example we have for instance
 \else

A \emph{run} in a weighted automaton is a path in the graph from a
state to a state. The run's weight is the product of all involved
transition weights   \emph{and} the last state's weight.
%More formally, let $A$ be the $10\times
%10$ matrix defining the weights of the automaton's transition,  and
%$v_0$ the vector of state weights, then we have the initial value
%problem $\dot\xx = A\xx$ with $\xx(0)=v_0$.
%
In the standard linear-time semantics of weighted automata, each
state  $\alpha$ is assigned a function $\sigma_\alpha: \mathbb{N}\rightarrow
\field$ (a \emph{stream}, in the terminology of Rutten
\cite{Rutten}) such that $\sigma_\alpha(i)$ is obtained by summing up all
the weights of the runs involving $i$ transitions   starting from
$s$; e.g. $\sigma_{x_1}(3)=1\times  \frac 2 3 \times \frac 3 4 + 1\times  \frac 1 3 \times \frac 3 2 = 1$. Standard results in
coalgebra (or a simple direct proof) ensure that this semantics is
in agreement with bisimulation, in the sense that for any pair of states
 (in our case, monomials) $\alpha$ and $\beta$,  $\alpha\sim_\ivp \beta$ if
and only $\sigma_\alpha=\sigma_{\beta}$; we refer the interested reader to
\cite{Rutten,IC12} for details on this construction and result.

When applied to our example, these results imply for instance\fi
%that linear time theorem of Subsection \ref{sub:weighted} implies
 \ that $x_1(t)= x_5(t)$. In fact, when invoked with this system and $\pi=\sum_{i=1}^{10} a_i x_i$ as inputs,
the double chain algorithm terminates at $m=2$ (in about 0.3 s;
this being  a linear system,   Gr\"{o}bner bases are never actually
needed), returning
\iflncs $\\ \pi[V_2] = \left( a_1 (x_6 - x_7) + a_2(x_8 -  x_9) +
a_3(x_6-x_2) + a_4(x_5- x_1  ) + a_5(\frac 3 2 x_8 - x_4  )\right.$\\
$\left.  +a_6(\frac 3 4 x_8- x_3)\right)[\field^6] $. \else
$\pi[V_2] = a_1 (x_6 - x_7) + a_2(x_8 -  x_9) +
a_3(x_6-x_2) + a_4(x_5- x_1  ) + a_5(\frac 3 2 x_8 - x_4  ) +
a_6(\frac 3 4 x_8- x_3) $.\fi
% represented by the
%following 6-parameters template {\small
%\begin{equation*}
%\pi[V_2] & = &
%a_1(2x_3 -  x_4) + a_2(\frac 4 3x_3 -  x_8) + a_3(
%x_7-x_2) +  a_4(  x_5-x_1) + a_5(x_7-x_6    ) + a_6(\frac 4 3 x_3 -
%x_9)\,.
%\end{equation*}
%}
\iflncs This result implies the expected   $x_1= x_5$, as well as
other equivalences, and a 60\% reduction in the minimal system.
\else\
 This implies the expected  equality $x_1= x_5$ (let $a_4=1$ and  $a_i=0$ for
$i\neq 4$ in the returned template), as well as other equalities,
such as $x_2=x_6=x_7$ and $x_8=x_9$. All in all, $V_2$ being  a 6-dimensional space, we
will have a 4-dimensional $W=V_2^\bot$, that is  a minimal system
with 4 equations,    a 60\% reduction.
\fi

%\vspm
\subsection{Example 2: nonlinear conservation laws}\label{sub:pendulum}
\begin{wrapfigure}{r}{0.18\textwidth}
\vspace*{-.35cm}
\usetikzlibrary{calc,patterns,angles,quotes}
  \begin{tikzpicture}%{0.2\textwidth}
    \coordinate (origo) at (0,0);
    \coordinate (pivot) at (1,5);
    \coordinate (roof) at (1,0);
    % draw axes
    \fill[black] (origo) circle (0.05);
    \draw[thick,gray,-] (origo) -- ++(1.6,0) node[black,right] {};
    \draw[thick,gray,-] (origo) -- ++(0,-2) node (mary) [black,below] {};
    % draw roof
    \fill[pattern = north east lines] ($ (origo) + (-1,0) $) rectangle ($ (origo) + (1,0.3) $);
    \draw[thick]  ($ (origo) + (-1,0) $) -- ($ (origo) + (1,0) $);
    \draw[thick] (origo) -- ++(300:2) coordinate (bob);
    \fill (bob) circle (0.2);
    \pic [thick,draw, <-, "$\theta$", angle eccentricity=1.5] {angle = bob--origo--roof};
  \end{tikzpicture}
  \vspace*{-.3cm}
\end{wrapfigure}

The law of the simple pendulum   is $\frac {d^2}{d t^2} \theta =
\frac g {\ell} \cos\theta$, where $\theta$ is the angle from
the roof to the rod measured clockwise, $\ell$ is the
length of the rod and $g$ is
  gravity acceleration (see picture on the right).
If we assume the initial condition
  $\theta(0)=0$, this
   can be translated into the   polynomial  initial value problem below, where
$\xx=(\theta,\omega,x,y)^T$. The meaning of the variables is
 $\omega=\dot\theta$,   $x=\cos\theta$   and
$y=\sin\theta$.
We assume for simplicity  $\ell = 1$ and $g=9$.
%
%\begin{wrapfigure}{r}{0.23\textwidth}
%  \vspace*{-1cm}
%\small
\[
\left\{
\begin{array}{rcl}
\dot{\theta}  & = & \omega\\
\dot{\omega} & = &  \frac g {\ell} x\\
\dot x & = & -y\omega\\
\dot y & = & x\omega\\
\xx(0) & = & (0,0,\ell,0)^T\,.
\end{array}\right.
\]
%\vspace*{.1cm}
%\centering
%\vspace*{-.8cm}
%\includegraphics[width=0.4\textwidth]{wa.eps}
%\vspace*{-9.4cm}
%\caption{\label{fig:plotx}This is a figure caption.}
%\vspace*{-1.3cm}
%\end{wrapfigure}
%\begin{wrapfigure}{r}{0.2\textwidth}
%
%\end{wrapfigure}
%
\noindent
For this system, the double chain algorithm  reports that there is
no nontrivial linear conservation law (after $m=6$ iterations and
about 0.3 s). We  then ask the
 algorithm to find all the conservation laws of order   two, %or less,
that is  we use the template ($\alpha$ ranges over monomials)
$\pi=\sum_{\alpha_i\,:\,\deg(\alpha_i)\leq 2} a_i\alpha_i$ as input.
The algorithm terminates after $m=16$ iterations (in about 7 s). The invariant  $J_{16}$
contains all the wanted conservation laws. The returned Gr\"{o}bner basis
   for it
  is $G=\{x^2 + y^2 - 1, \omega^2 - 18y\}$. The first term here just
  expresses the trigonometric identity
  $(\cos\theta)^2+(\sin\theta)^2=1$. Recalling that the (tangential) speed of
  the bob is $v=\ell\dot\theta=\ell\omega$, and that its vertical distance  from the roof is
  $h=\ell \sin\theta = \ell y$, we see that the second term,
  considering our numerical values
  for $\ell,g$, is equivalent to the equation
   $\frac 1 2 v^2 = g h$, which, when multiplied by the mass $m$ of the bob, yields the law of conservation
   of energy   $\frac 1 2 mv^2 = mg h$ (acquired kinetic energy = lost potential energy).

\iflncs\section{Future and related work}\label{sec:concl}\vspm
\else
\section{Conclusion, further and related work}\label{sec:concl}
\fi

\iflncs We briefly outline future work and   related work below,
referring the reader to \cite{Full} for a more comprehensive
discussion. \else We have presented a framework for automatic
reasoning and reduction in systems of polynomial \ode's.   In
particular, we offer algorithms to: (1) compute the most general set
of identities valid in the system that fit a user-specified
template; and, (2)
  build  a minimal system equivalent to the original
one. These algorithms are based on a mix of simple algebraic and
coalgebraic techniques.
\fi

\subsection{Directions for further work}
%\paragraph{Directions for future work}
%There are several directions along which the present work can be
%extended.
Scalability of our approach is an issue, as, already for simple systems,
the Gr\"{o}bner basis construction involved in the double chain algorithm
can be  computationally quite demanding. Further experimentation,
relying on a well-engineered implementation of the method, and considering sizeable
case studies, is called for in order to assess  this aspect.
%Another limitation of the present approach is its dependence on fixed initial values for the problem.
\iflncs\else One would also like to extend the present  approach so
as to deal with \emph{regions} of possible initial values, rather
than fixing one such value. This is important, e.g., in the
treatment of hybrid systems (see below). After the short version of the present paper \cite{Bor17} appeared, some preliminary progress towards this goal has been made, see \cite{Bor18}. Concerning minimization, we
note that the reduced system may not preserve the structure of the
original one, e.g. as to the meaning of variables: this may be problematic in
certain application domains, such as system biology. In the future,
we intend to investigate this issue. \fi Approximate reductions in
the sense of System Theory \cite{Antoulas} are also worth
investigating. One problem of the approach described in Section \ref{sec:approx} is that the reduced system depends on   a fixed initial condition  $v_0$.
Obtaining   bounds on the approximation error is another aspect that deserves further investigation.
Some preliminary progress on these issues is reported in~\cite{HSCC18}. % linearization.
%Approximation and linearization techniques
%Extensions of the theory allowing us to deal  with \emph{parameters}
%in the equations would also be worth considering.

\subsection{Related work}
Bisimulations for weighted automata are related to our approach,
because, as argued in subsection \ref{sub:linear},  Lie-derivation
can be naturally represented by  such an automaton. Algorithms for
computing largest bisimulations on \emph{finite} weighted automata
have been studied by Boreale et al. \cite{Bor09,IC12}. A crucial
ingredient in these algorithms is the representation of
bisimulations as   finite-dimensional vector spaces. Approximate
versions of this technique have also been recently considered in
relation to Markov chains \cite{Bor15}. As discussed in Remark
\ref{rem:linsys}, in the case of linear systems, the algorithm in
the present paper reduces to that of \cite{Bor09,IC12}.
%\iflncs\else
Algebraically, moving from linear to polynomial systems corresponds
to  moving from vector spaces to ideals, hence from linear bases to
Gr\"{o}bner bases. From the point of view automata, this step leads
to considering infinite weighted automata. In this respect, the
present work may be also be related to the automata-theoretic
treatment of linear \ode's   by Fliess and Reutenauer \cite{FR83}.
%\fi
  %corresponds, as seen in this
%paper,

%\paragraph{Relation with the work of Cardelli et al. \cite{MircoPopl}}
Although there exists
 a rich literature dealing with linear
  aggregation
   %and approximation
   of systems of \ode's (e.g. \cite{Antoulas,LiRabToth,Toth,Okino}),
   we are not aware of  fully automated approaches to
   %reasoning and
   minimization (Theorem  \iflncs\ref{th:exactMinimal}\else \ref{th:minimal}\fi),
 with the notable exception of a series of  recent works by Cardelli and
    collaborators \cite{MircoPopl,DiffEq1,DiffEq2}. Mostly related to ours is
    \cite{MircoPopl}.  There,
     for  an extension of the polynomial \ode\  format called \textsf{IDOL}, the authors introduce two flavours
      of \emph{differential equivalence}, called \emph{Forward} (\fde) and \emph{Backward}
      (\bde). They provide a symbolic,
        SMT-based partition refining algorithms to compute the largest equivalence of each type.
        \iflncs While \fde\ is unrelated with our equivalence, \bde\ can be compared directly to our  \lbisim-bisimulation.
        \else
       \fde\ groups variables   in such a way that
        the corresponding quotient system recovers the \emph{sum}
        of the original solutions in each class, whatever the initial condition.
        However, precise information on the individual original solutions cannot in general
         be recovered from the reduced system. In \bde, variables grouped together are guaranteed to
         have  the same solution.
        Therefore the quotient system permits in this case to fully recover  the original solutions.
         As such, \bde\ can be compared directly to our  \lbisim-bisimulation.
%There are some important differences from our approach.

\fi
An important difference   is
that \bde\  may  tell apart  variables that have
the same solution. As already seen, this is not the case with \lbisim-bisimilarity,
which  is  correct and complete.
%In fact, in a very
%precise sense  (discussed in Section \ref{sec:examples}), \bde\ is
%branching-time, whereas \lbisim-bisimilarity is linear-time.
An important consequence of this difference is that the quotient
system produced by \bde\ is not minimal, whereas that produced by
\lbisim-bisimulation is,  in a precise sense. In concrete cases, this
 may imply a significant size difference. For example, in the linear
  system of Subsection  \ref{sub:linear},
\bde\ finds only two equalities\footnote{Which are  $x_6=x_7$ and $x_8=x_9$, as checked with the Erode tool by the  same authors \cite{Erode}.}, leading to a quotient system of eight states;
   on the other hand,  the minimal system produced by \lbisim-bisimulation has four states.
For what concerns
reasoning, we note that, being based on partitions of variables,
 \bde\ cannot  express relations involving polynomial, or
 even linear,
combinations of variables. \fi
Finally, the approach of
\cite{MircoPopl} and ours rely on two quite different algorithmic
decision techniques, SMT and Gr\"{o}bner bases, both of which have
exponential worst-case complexity. As shown by the experiments reported  in \cite{MircoPopl},
 in practice \bde\ and \fde\ have proven   quite effective at system reduction. At the moment,
 we lack similar experimental evidence for \lbisim-bisimilarity.
 \iflncs\else
  We also   note that, limited to the case of polynomials of degree two, a polynomial algorithm for
  \bde\ exists \cite{DiffEq1}. %Furthermore, we note that Further experimentation might
%clarify in what concrete domains of applications one approach can be
%more effective than the other.
\fi

\iflncs\else
Linear aggregation  and lumping of (polynomial) systems of \ode's
are well known in the literature, se e.g. \cite{Antoulas,Okino,LiRabToth,Toth}
and references therein. However, as pointed out by Cardelli et al.
\cite{MircoPopl}, no general algorithms for computing the largest
equivalence, hence the minimal \emph{exact} reduction (in the sense
of our Theorem \ref{th:minimal})
 was known.
 \fi

The seminal paper of Sankaranarayanan, Sipma  and Manna \cite{San04}
introduced polynomial ideals to find invariants of hybrid systems.
%, of which polynomial \ode's can
%be seen as a subclass.
Indeed,  the study of the safety of hybrid systems can be shown to
reduce constructively to the problem of generating invariants for
their differential equations \cite{Pla12}. The results   in
\cite{San04} have been subsequently refined and simplified by
Sankaranarayanan using \emph{pseudoideals} \cite{San10}, which
enable the discovery of polynomial invariants of a special form.
Other authors have adapted this approach to the case of imperative
programs, see e.g. \cite{Farewell,Muller,Carbonell} and
references therein. %In all these cases, though, the proposed
%technique can either only find   invariants of interest, in a spe,
%or find all of them only at cost of more or less severe restriction
%on the source language \cite{FindAll}.
Reduction and minimization seem to be not  a concern in this field.

%\ode's play of course a crucial role also in hybrid systems, which exhibit a mix of continuous and discrete transitions.
\iflncs Still in the field of formal verification of hybrid systems,
mostly related to ours is  Ghorbal and Platzer's recent work  on
polynomial invariants \cite{Pla14}. \else Platzer  has introduced
\emph{differential dynamic logic} to reason on hybrid systems
\cite{DDL}. The rules of this logic implement a fundamentally
inductive, rather than coinductive, proof method. Mostly related to
ours is  Ghorbal and Platzer's recent work  on polynomial invariants
\cite{Pla14}. \fi
 One one hand, they characterize
algebraically invariant regions  of   vector fields -- as opposed to
initial value problems, as we do. On the other hand, they offer
sufficient conditions under which the trajectories induced by
specific initial values  satisfy  all instances of a polynomial
template (cf. \cite[Prop.3]{Pla14}). The latter result compares with
ours,  but  the resulting method appears to be not (relatively)
complete in the sense of our double chain algorithm. Moreover, the
computational prerequisites of \cite{Pla14} (symbolic linear
programming, exponential size matrices, symbolic root extraction)
are very different from ours, and   much more demanding. Again,
minimization   is not addressed.

Ideas from Algebraic Geometry have been fruitfully applied also in Program Analysis. Relevant to our work is M\"{u}ller-Olm and Seidl's \cite{Muller}, where  an algorithm to compute all   polynomial invariants up to a given degree of an imperative program  is provided. Similarly to what we do, they  reduce the core   problem to a linear algebraic one. However, being the setting in \cite{Muller} discrete rather than continuous, the  techniques employed there are otherwise quite different, mainly because: (a) the construction of the ideal chain is driven by the program's operational semantics, rather than by Lie derivatives; (b)   the found polynomial invariants  must be valid under \emph{all} initial program states, not  just under the  user specified one. If transferred  to  a continuous setting,   condition (b) would  lead in most cases to  trivial invariants.

%\ifmai
In nonlinear  Control Theory, there is a huge amount of literature
on \emph{Model Order Reduction} (\mor), that aims at reducing the
size of a given system, while preserving some properties of
interest, such as stability and passivity. A well established
approach   relies on building truncated Taylor  expansions of the
given sysytems \cite{Pileggi,Phillips}, repeated at various points
along a trajectory of interest, to   keep the approximation error
globally small: a technique known as \emph{trajectory piece-wise
linear} ({\sc tpwl}) \mor, see e.g. \cite{Rewi}. One wonders whether
our approximate  linearization technique of Section \ref{sec:approx}
might conveniently serve as a   building block of this strategy.
%\fi

The present paper is the extended and revised version of \cite{Bor17}. W.r.t. \cite{Bor17}, here we include complete proofs, the discussion of up-to techniques in Section \ref{sec:coalg}, the approximate linearization  technique of Section \ref{sec:approx} and an extended and updated discussion of further and related work in the present section.

%\vspm
%\vspace*{.3cm}
%\noindent
\section*{Acknowledgments} The author has benefited from stimulating discussions with Mirco
Tribastone. Two anonymous \textsc{lmcs} reviewers have provided
valuable comments that have helped to improve the presentation.

%\bibliographystyle{abbrvnat}

% The bibliography should be embedded for final submission.

%\newpage
\appendix
\section{Proofs and additional technical material}\label{app:proofs}

\subsection{Proofs}\label{sub:proofs}
\noindent
\begin{proof_of}{Theorem \ref{th:bisim}}
We check each of the equivalences (\ref{eq:c0}--\ref{eq:c2}) in
turn.
\begin{itemize}
\item If $p\sim_\ivp q$ then $p-q\in \ker(\sim_\ivp)$ by definition of $\ker$.
Conversely, assume $p-q\in   {\ker(\sim_\ivp)  } $ $\subseteq $ $\ide{\ker(\sim_\ivp)}$. Since,  by first part of Lemma  \ref{lemma:prefixp} with $R=\sim_\ivp$,
 the
last set is an invariant, by the second part of the same
lemma we obtain the wanted $p\sim_\ivp q$, hence \eqref{eq:c0}.
%we first prove that \mbox{$\ker(\sim_\ivp)$} is an invariant. Indeed,
%since $\sim_\ivp$ in $C_\ivp$ corresponds to equality in $\A$
%(Theorem \ref{th:finality}), and equality is of course preserved by sum
%and product in $\A$, we easily obtain that
%$\ker(\sim_\ivp)=\ide{\ker(\sim_\ivp)}$, hence $\ker(\sim_\ivp)$ is an invariant by the first part of Lemma  \ref{lemma:prefixp}.
%From this fact,  letting
%$R=\sim_\ivp$ in the second part of Lemma \ref{lemma:prefixp}, the wanted implication ($p-q\in \ker(\sim_\ivp)$ implies $p\sim_\ivp q$),
% hence
%\eqref{eq:c0}, follows.

\item Since $p\sim_\ivp q$ implies $(p-q)(t)=0$, and $r(t)=0$ implies
$r\sim_\ivp 0$ (in both cases   by Theorem \ref{th:finality}), from
the definition of $\ker(\sim_\ivp)$  the second equality
\eqref{eq:c1} immediately follows.

\item A polynomial behaviour $p(\xx(t))$ is identically 0 in $\A$ if and
only if all its derivatives $\frac d{dt^j} p(\xx(t))$, $j\geq 0$,
vanish at 0, and this, via \eqref{eq:liederj} and
\eqref{eq:initial}, yields \eqref{eq:c3}.

\item Note that if $p\in I$,  with $I$ an invariant, then
$p^{(j)}(v_0)=0$ for each $j\geq 0$ (easily shown by induction on
$j$). Conversely, if  $p^{(j)}(v_0)=0$ for each $j$, then
$J=\ide{\{p,p^{(1)}, p^{(2)},...\}}$ is an invariant and contains
$p$. To check invariance of $J$,  consider a generic $q\in J$,
$q=\sum_i h_i p^{(j_i)}$: it is immediate  that $q(v_0)=0$ and that
$\lie(q)= \sum_i \lie(h_i) p^{(j_i)} + \sum_i h_i p^{(j_i+1)}\in J$.
This way we have also proven the last equation \eqref{eq:c2}.
\qedhere
\end{itemize}
\end{proof_of}

\noindent
%\begin{lem}[Lemma \ref{lemma:stab}]
%Let $V_m,J_m$ be the sets returned by the algorithm. Then %sequence  of sets $V_i,J_i$ stabilizes at $m$:  that is,
%for each $j\geq 1$, one has
% $V_m=V_{m+j}$ and $J_m=J_{m+j}$.
%\end{lem}
\begin{proof_of}{Lemma \ref{lemma:stab}}
We proceed by induction on $j$. The  base case $j=1$ follows from
the definition of $m$. Assuming by induction hypothesis that
$V_m=\cdots=V_{m+j}$ and that $J_m=\cdots=J_{m+j}$, we prove now
that
  $V_m=V_{m+j+1}$ and that $J_m=J_{m+1+1}$. The key to the proof
is the following fact
\begin{eqnarray}
\pi^{(m+j+1)}[v]& \in &J_m  \;\; \text{ for each }v\in
V_m\,.\label{rel:fund}
\end{eqnarray}
From this fact the thesis will follow, indeed:
\begin{enumerate}
\item $V_m=V_{m+j+1}$. To see this, observe that for each $v\in V_{m+j}=V_m$
 (the equality here follows from the induction hypothesis), it follows from  \eqref{rel:fund} that
$\pi^{(m+j+1)}[v]$ can be written as a finite sum of the form
$\sum_{l} h_l\cdot \pi^{(j_l)}[u_{l}]$, with $0\leq j_l\leq m$ and
$u_l\in V_m$. As a consequence, $\pi^{(m+j+1)}[v](v_0)=0$, which
shows that $v\in V_{m+j+1}$. This   proves that $V_{m+j+1}\supseteq
V_{m+j}=V_m$; the reverse inclusion is obvious;
\item $J_m=J_{m+j+1}$. As a consequence of $V_{m+j+1}=V_{m+j}(=V_m)$ (the previous point), we can write
\begin{eqnarray*}
J_{m+j+1} & = &\ide{ \cup_{i=0}^{m+j}\pi^{(i)}[V_{m+j}] \cup
\pi^{(m+j+1)}[V_{m+j}]}\\
           & = & \ide{ J_{m+j} \cup
\pi^{(m+j+1)}[V_{m+j}]}\\
           & = & \ide{ J_{m} \cup
\pi^{(m+j+1)}[V_{m}]}
\end{eqnarray*}
where the last step follows by induction hypothesis. From
\eqref{rel:fund}, we have that \linebreak $\pi^{(m+j+1)}[V_{m}]\subseteq J_m$,
which implies the thesis for this case, as $\ide{J_m}=J_m$.
\end{enumerate}
We prove now \eqref{rel:fund}. Fix any $v\in V_m$. First, note that
$\pi^{(m+j+1)}[v]= \lie(\pi^{(m+j)}[v])$ (here we are using
\eqref{eq:templder}). As by induction hypothesis
$\pi^{(m+j)}[V_m]=\pi^{(m+j)}[V_{m+j}]\subseteq J_{m+j}=J_m$, we
have that $\pi^{(m+j)}[v]$ can be written as a finite sum $\sum_{l}
h_l\cdot \pi^{(j_l)}[u_{l}]$, with $0\leq j_l\leq m$ and $u_l\in
V_m$. Applying the rules of Lie derivatives \eqref{eq:liesum},
\eqref{eq:lieprod}, we find  that $\pi^{(m+j+1)}[v] =
\lie(\pi^{(m+j)}[v])$ equals
\begin{equation*}
%\pi^{(m+j+1)}[v]\; = \; \lie(\pi^{(m+j)}[v]) & = &
\sum_{l}\left( h_l\cdot
\pi^{(j_l+1)}[u_{l}]+ \lie(h_l)\cdot \pi^{(j_l)}[u_{l}]\right)\,.
\end{equation*}
Now, for  each $u_l$, $u_l\in V_m=V_{m+1}$,  each term
$\pi^{(j_l+1)}[u_{l}]$, with $0\leq j_l+1\leq m+1$, is by definition
in $J_{m+1}=J_m$. This shows that $\pi^{(m+j+1)}[v]\in J_m$, as
required.
\end{proof_of}

%\begin{cor}[Corollary \ref{cor:redbisim}]
%Let $p,q\in\field[\xx]$. Then $p\sim_{F} q$ in $C_F$ if and only if
%$p(B\yy) \sim_{G} q(B\yy)$ in $C_G$. In particular, $x_i\sim_F x_j$
%in if and only if row  $i$ equals row $j$ in $B$.
%\end{cor}
\noindent
\begin{proof_of}{Corollary \ref{cor:redbisim}}
Concerning the first part, let us denote by $\mathcal{Z}_\redpr$ the largest
invariant induced by $\redpr$ in the polynomial ring $\field[\yy]$,
according to Theorem \ref{th:bisim}. We have: $p\sim_{\ivp} q$ in
$\field[\xx]$ if and only if $(p-q)\in \Zzero$ (Theorem
\ref{th:bisim}\eqref{eq:c1}) if and only if $(p-q)\circ \xx(t)=0$
(Theorem \ref{th:bisim}\eqref{eq:c0}) if and only if $(p-q)\circ
B\yy(t)=0$ (Theorem \ref{th:exred}) if and only if
$(p(B\yy)-q(B\yy))\left(t\right)=0$ if and only if
$(p(B\yy)-q(B\yy))\in  \mathcal{Z}_\redpr  $ (Theorem
\ref{th:bisim}\eqref{eq:c0}) if and only if $p(B\yy)\sim_{\redpr}
q(B\yy)$ in $\field[\yy]$ (Theorem \ref{th:bisim}\eqref{eq:c1}).

Concerning the second part, denoting by $e_i$ the $i$-th canonical
vector in $\field^N$,
 note that: $x_i\sim_\ivp x_j$ if and only if $x_i-x_j\in \Zzero$ if and only if $e_i-e_j\in V_m$
 if and only if $e_i-e_j\bot W$ if and only if $B^T(e_i-e_j)=0$, from which the thesis follows for this case.
\end{proof_of}

\subsection{Pseudoideals}\label{sub:pseudo}
We briefly discuss a sufficient condition for  establishing the
condition \eqref{eq:stabilize}, that is $J_{i+1}=J_i$, which does
not involve   Gr\"{o}bner bases, but only linear algebraic
computations, and can therefore lead to a gain in efficiency. We ill
make use of a bit of new notation. For a set of polynomials $S$ and
an integer $k\geq 0$,  denote by $\ide{S}_k$ the  subset of $\ide S$
generated from $S$ by only using multiplier polynomials $h_j$ of
degree $\leq k$ (cf. equation \eqref{eq:ide}); this is a
\emph{pseudo ideal of degree $k$}, in the terminology of Col\'{o}n
\cite{Col}. We can choose     $k\geq 0$ and replace
\eqref{eq:stabilize} by the following stronger    condition
\begin{eqnarray}
\pi^{(i+1)}[B_i]&\subseteq & \ide{\cup_{j=0}^i
\pi^{(j)}[B_i]}_k\,.\label{eq:stabilizek}
\end{eqnarray}
If \eqref{eq:stabilizek} is true then of course  also
\eqref{eq:stabilize} is true, while the converse is not valid in
general. Condition \eqref{eq:stabilizek} can be checked by linear
algebraic techniques, which do not involve Gr\"{o}bner bases
computations. Indeed, a pseudo ideal $\ide{ S}_k$ has the structure
of  a vector space over $\field$ of dimension $|S|\cdot M$, where
$M$ is the number of distinct monomials of degree $\leq k$. However,
the resulting algorithm is not guaranteed to terminate.

\subsection{Building an orthonormal basis of $V^\bot$}\label{sub:basis}{%\em
We use here to the terminology of Section \ref{sec:minim}. Let us  first work out a convenient characterization of  the space $W=V^\bot$. %, the
%orthogonal complement of $V$.
Consider the successive Lie derivatives of the vector
$\xx=(x_1,...,x_N)^T$, taken componentwise, that is the vectors of
polynomials $\xx^{(j)}=(x_1^{(j)},...,x_N^{(j)})^T$, for $j=0,1,..$.
Once evaluated at   $v_0$, these become vectors in $\field^N$. We
claim that
\begin{eqnarray}
W & =
&\myspan\left\{\xx(v_0),\xx^{(1)}(v_0),...,\xx^{(m)}(v_0)\right\}\,.\label{eq:W}
\end{eqnarray}
%
%Now, denoting by $\langle\cdot,\cdot\rangle$ the scalar product
%between vectors, and by $(\cdot)^\bot$ the orthogonal complement
%operation over subspaces of $\field^{N}$,
Indeed, we have, by definition of $V$ (here $v=(\lambda_1,...,\lambda_N)^T$
denotes a generic vector in $\field^N$)
\begin{eqnarray*}
V & = & V_m\\
  & = & \left\{v\,:\,\sum_{i=1}^N \lambda_i x^{(j)}_i(v_0)=0 \text{ for
  }j=0,...,m\right\}\\
  & = & \left\{v\,:\,\langle v,\xx^{(j)}(v_0)\rangle =0 \text{ for
  }j=0,...,m\right\}\\
  & = & \left\{\xx(v_0),\xx^{(1)}(v_0),...,\xx^{(m)}(v_0)\right\}^\bot
%  & = & W^{\bot}
\end{eqnarray*}
which is equivalent to \eqref{eq:W}. Note that $l=\dim(W)\leq m+1, N$
(and typically one will have $l\ll N$). Therefore, a way to build an
orthonormal basis $B$ for $W$ is just to apply the Gram-Schmidt
orthonormalization process to the set of vectors on the right-hand
side of \eqref{eq:W}. This process can in fact carried out
incrementally, as  the vectors of Lie derivatives $\xx^{(j)}$ are
computed. %\cite{smith02}
}
%\end{rem}


\begin{thebibliography}{10}
%\softraggedright

\bibitem{Antoulas} A.C. Antoulas.
\emph{Approximation of Large-scale Dynamical Systems}. SIAM, 2005.

\bibitem{PL} V.I. Arnold.  \emph{Ordinary Differential Equations}.  The MIT
Press, ISBN 0-262-51018-9, 1978.


\bibitem{Arnoldi}
W. E. Arnoldi. The principle of minimized iterations in the solution
of the matrix eigenvalue problem.  \emph{Quarterly of Applied
Mathematics}, vol. 9, pp. 17-29, 1951.


\bibitem{CTMC} M. Bernardo.
A survey of Markovian behavioral equivalences. In \emph{Formal
Methods for Performance Evaluation}, vol. 4486 of LNCS, pages
180-219. Springer, 2007.

\bibitem{RifBio2} M. L. Blinov, J. R. Faeder, B. Goldstein, and W. S. Hlavacek.
BioNet-Gen: software for rule-based modeling of signal transduction
based on the interactions of molecular domains.
\emph{Bioinformatics}, 20(17): 3289-3291, 2004.

\bibitem{IC12} F. Bonchi, M.M. Bonsangue, M. Boreale, J.J.M.M.
Rutten, and A. Silva. A coalgebraic perspective on linear weighted
automata. \emph{Inf. Comput.} 211: 77-105, 2012.


\bibitem{Bor09} M. Boreale. Weighted Bisimulation in Linear Algebraic Form.
 \emph{Proc. of CONCUR 2009}, LNCS 5710,  pp. 163-177, Springer, 2009.

\bibitem{Bor15}
M. Boreale. Analysis of Probabilistic Systems via Generating
Functions and Pad\'{e} Approximation. \emph{ICALP 2015} (2):
82-94, LNCS 9135, Springer, 2015. Extended version available as
DiSIA working paper 2016/10,
\url{http://local.disia.unifi.it/wp_disia/2016/wp_disia_2016_10.pdf}.

\bibitem{Bor17} M. Boreale. Algebra, coalgebra, and minimization in
polynomial differential equations. In {\em Proc. of FoSSACS 2017}:71--87, LNCS 10203, Springer, 2017.
Conference version of the present paper.

\bibitem{Bor18} M. Boreale.  Complete algorithms for algebraic strongest postconditions  and
weakest preconditions in  polynomial \ode's. In \emph{Proc. of
SOFSEM'18}, LNCS 10706: 442--455, Springer. Full version available as
%\emph{CoRR, abs/1708.05377},
\url{http://arxiv.org/abs/1708.05377},
2017.

\bibitem{HSCC18} M. Boreale.  Algorithms for exact and approximate linear abstractions of polynomial
 continuous systems. \emph{HSCC 2018}: 207--216, ACM, 2018.
% In \emph{Proceedings of the 21st International Conference on Hybrid Systems:
%              Computation and Control (part of {CPS} Week), {HSCC} 2018}, Porto,
%               Portugal, April 11-13, 2018, pp. 207--216, doi 10.1145/3178126.3178137, ACM, 2018.

\bibitem{Farewell}
D. Cachera, Th. Jensen, A. Jobi, and F. Kirchner. Inference of
Polynomial Invariants for Imperative Programs: A Farewell to
Gr\"{o}bner Bases. \emph{SAS 2012}, LNCS 7460: 58-74, Springer,
2012.


\bibitem{RifBio3} L. Cardelli. On process rate semantics. \emph{Theoretical Computer
Science}, 391(3):190-215, 2008.

\bibitem{MircoPopl} L. Cardelli, M. Tribastone, M. Tschaikowski, and A. Vandin.
Symbolic Computation of Differential Equivalences, \emph{POPL 2016}, ACM, 2016.

\bibitem{DiffEq1} L. Cardelli, M. Tribastone, M. Tschaikowski, and A. Vandin.
Efficient Syntax-driven Lumping of Differential Equations, \emph{TACAS
2016}:93-111, LNCS 9636, Springer, 2016.

\bibitem{DiffEq2} L. Cardelli, M. Tribastone, M. Tschaikowski, and A. Vandin.
Comparing Chemical Reaction Networks: A Categorical and Algorithmic
Perspective, \emph{LICS 2016}, IEEE, 2016.

\bibitem{Erode}
L. Cardelli, M. Tribastone, M. Tschaikowski, and A. Vandin. ERODE:
Evaluation and Reduction of Ordinary Differential Equations.
Available from \url{http://sysma.imtlucca.it/tools/erode/}.

\bibitem{RifBio4} F. Ciocchetta and J. Hillston. Bio-PEPA:A framework for the modelling and
analysis of biological systems. \emph{Theoretical Computer Science},
410 (33-34):3065-3084, 2009


\bibitem{Col} M. Col\'{o}n.  Polynomial approximations of the relational semantics of
imperative programs. \emph{Science of Computer Programming} 64:
76-9, 2007.

\bibitem{Cox} D. Cox, J. Little, and D. O'Shea. \emph{Ideals, Varieties, and Algorithms An Introduction to Computational
Algebraic Geometry and Commutative Algebra}. Undergraduate Texts in
Mathematics, Springer, 2007.


\bibitem{FR83} M. Fliess,  and C. Reutenauer.
Theorie de Picard-Vessiot des Syst\`emes Reguliers. \emph{Colloque
Nat. CNRS-RCP567, Belle-ile  sept. 1982}, in \emph{Outils et
Mod\`eles Math\'ematiques pour l'Automatique l'Analyse des
syst\`emes et le tratement du signal}. CNRS, 1983.

\bibitem{covariety}
H.P. Gumm, Tobias Schr\"{o}der. Covarieties and complete
covarieties. \emph{Theoretical Computer Science} 260(1):71-86,
Elsevier, 2001.


\bibitem{Pla14}
K. Ghorbal, A. Platzer. Characterizing Algebraic Invariants by
Differential Radical Invariants. \emph{TACAS 2014}: 279-294, 2014.
Extended version available from
\url{http://reports-archive.adm.cs.cmu.edu/anon/2013/CMU-CS-13-129.pdf}.


\bibitem{Pileggi} P. Li and L. Pileggi. Compact reduced-order
modeling of weakly nonlinear analog and RF circuits. \emph{ IEEE
Trans. Comput.-Aided Des. Integr. Circuits Syst.}, vol. 24, no. 2,
pp. 184-203, 2005.


\bibitem{LiRabToth}
 G. Li, H. Rabitz, and J. T\'{o}th. A general analysis of exact nonlinear lumping
in chemical kinetics. \emph{Chemical Engineering Science} 49 (3),
343-361, 1994.

\bibitem{Muller}
M. M\"{u}ller-Olm and H. Seidl.  Computing polynomial program
invariants. \emph{Information Processing Letters} 91(5), 233-244,
2004.

%\bibitem{Norris} J. Norris. \emph{Markov Chains}. Cambridge Series in Statistical and
%Probabilistic Mathematics. Cambridge University Press, 1998.

\bibitem{Okino}
M.S. Okino and M.L. Mavrovouniotis. Simplification of mathematical
models of chemical reaction systems. \emph{Chemical Reviews},
2(98):391-408, 1998.

\bibitem{Phillips}
J.R. Phillips. Projection-based approaches for model reduction of
weakly nonlinear time-varying systems. \emph{ IEEE Trans.
Comput.-Aided Des.}, vol. 22, no. 2, pp. 171-187, 2003.

\bibitem{DDL}
A. Platzer. Differential dynamic logic for hybrid systems. \emph{J.
Autom. Reasoning} 41(2), 143-189, 2008.

\bibitem{Pla12}
A. Platzer. Logics of dynamical systems. In  \emph{LICS 2012}:
13-24, IEEE, 2012.



\bibitem{Rewi}
M. Rewienski and J. White. A trajectory piecewise-linear approach to
model order reduction and fast simulation of nonlinear circuits and
micromachined devices. \emph{ IEEE Trans. Comput.-Aided Des.}, vol.
22, no. 2, pp. 155-170, 2003.


\bibitem{Carbonell}
E. Rodr\'{i}guez-Carbonell and D. Kapur. Generating all polynomial
invariants in simple loops. \emph{Journal of Symbolic Computation}
42(4), 443-476, 2007.


\bibitem{Roychowdhury}
J. Roychowdhury. Reduced-order modelling of time-varying systems.
\emph{IEEE Trans. Circuits Syst.-II: Analog Digital Signal
Process.}, vol. 46, no. 10, pp. 1273-1288, 1999.



\bibitem{Rutten} J.J.M.M. Rutten. Behavioural differential equations: a coinductive
calculus of streams, automata, and power series. \emph{Theoretical
Computer Science},   308(1--3): 1--53, 2003.

\bibitem{Saad}
Y. Saad. \emph{Iterative methods for sparse linear systems}. SIAM,
2003.


\bibitem{San} D. Sangiorgi. Beyond Bisimulation:
The ``up-to" Techniques. \emph{FMCO 2005}: 161-171, 2005.


\bibitem{San04} S. Sankaranarayanan, H. Sipma, and Z. Manna. Non-linear loop invariant
generation using Gr\"{o}bner bases. \emph{POPL 2004}.

\bibitem{San10} S. Sankaranarayanan. Automatic invariant generation for hybrid
systems using ideal fixed points. \emph{HSCC 2010}: 221-230, ACM, 2010.

\bibitem{Tiwa}
A. Tiwari.  Approximate  reachability  for  linear  systems.  \emph{
HSCC 2003}: 514-525, ACM, 2003.

\bibitem{Toth}
J. T\'{o}th, G. Li, H. Rabitz, and A. S. Tomlin. The effect of lumping
and expanding on kinetic differential equations. \emph{SIAM Journal
on Applied Mathematics}, 57(6):1531-1556, 1997.


\bibitem{Fluid} M. Tribastone, S. Gilmore, and J. Hillston.
Scalable differential analysis of process algebra models. \emph{IEEE
Trans. Software Eng.}, 38(1):205-219, 2012.


\bibitem{RifBio1} E. O. Voit. Biochemical systems theory: A review. \emph{ISRN Biomathematics}, 2013:53, 2013.


\end{thebibliography}
\end{document}